\documentclass[12pt]{article}
\usepackage{amsmath,mathtools,amsfonts,amsthm,setspace,xcolor,mathrsfs,amssymb,comment,enumitem, hyperref}
\usepackage[margin=1.25in]{geometry}
\usepackage{natbib}
\usepackage[textwidth=30mm]{todonotes}

\newcommand{\dra}{\rightarrow \hskip-8pt \rightarrow}
\newcommand{\re}{\mathbb{R}}
\newcommand{\na}{\mathbb{N}}
\newcommand{\N}{\mathbb{N}}

\newcommand{\id}{\mathbb{I}}

\newcommand{\BR}{\text{BR}}

\def\C{{\mathcal C}}
\def\D{{\mathcal D}}

\def\F{{\mathcal F}}
\def\G{{\mathcal G}}
\def\H{{\mathcal H}}
\def\B{{\mathcal B}}

\newcommand{\veps}{\varepsilon}

\newcommand{\Eqref}[1]{Eq.\,(\ref{#1})}

\font\eightrm=cmr8

\newtheorem{theorem}{Theorem}

\newtheorem{remark}[theorem]{Remark}

\newtheorem*{claim}{Claim}

\newtheorem{lemma}[theorem]{Lemma}
\newtheorem{definition}[theorem]{Definition}
\newtheorem{example}[theorem]{Example}

\begin{document}

\title{\vspace{-1.5cm}A general definition of perfect equilibrium}
\author{
        \sc J\'{a}nos Flesch
        \and
        \sc Christopher Kops
        \and
         \sc Dries Vermeulen
        \and
        \sc Anna Zseleva
        \thanks{Maastricht University, School of Business and Economics, Department of Quantitative Economics, P.O.\ Box 616, 6200 MD Maastricht, The Netherlands. Flesch: {\tt j.flesch@maastrichtuniversity.nl}; Kops: {\tt j.kops@maastrichtuniversity.nl}; Vermeulen: {\tt d.vermeulen@maastrichtuniversity.nl}, Zseleva: {\tt anna.zseleva@maastrichtuniversity.nl}. The authors would like to thank Philip Reny, Klaus Ritzberger, Dmitriy Kvasov, and the audiences at SING 2025, SAET 2025 Conference, and Maastricht MLSE seminar for very helpful discussions.}
}
\maketitle

\begin{abstract}
We propose a general definition of perfect equilibrium which is applicable to a wide class of games. A key feature is the concept of completely mixed nets of strategies, based on a more detailed notion of carrier of a strategy. Under standard topological conditions, this definition yields a nonempty and compact set of perfect equilibria. For finite action sets, our notion of perfect equilibrium coincides with \citeauthor{Selten: 1975}'s \citeyearpar{Selten: 1975} original notion. In the compact-continuous case, perfect equilibria are weak perfect equilibria in the sense of \cite{Simon Stinchcombe:1995}. In the finitely additive case, perfect equilibria in the sense of \cite{Marinacci: 1997} are perfect. Under mild conditions, perfect equilibrium meets game-theoretic desiderata such as limit undominatedness and invariance. We provide a variety of examples to motivate and illustrate our definition. Notably, examples include applications to games with discontinuous payoffs and games played with finitely additive strategies.
\vskip6pt


\end{abstract}

\clearpage
\section{Introduction}
Perfect equilibrium \citep{Selten: 1975} refines the concept of Nash equilibrium. A perfect equilibrium is a Nash equilibrium which is robust to opponents making small mistakes in their choice of equilibrium actions. In the finite action case, such mistakes take the form of playing every action with strictly positive probability. For infinite action sets, however, this interpretation may cease to be valid. For instance, when choosing actions from within the unit interval it is no longer possible to simultaneously play each individual action with strictly positive probability. This paper contributes to the line of research that generalizes the notion of perfection in a way that makes it applicable to a wide range of strategic form games with (possibly) infinite action sets.
\vskip6pt

Many economic interactions are modeled as games with infinite action spaces. Examples run from auctions \citep{Vickrey: 1961}, to oligopolistic competition \citep{Bertrand: 1883,Cournot: 1838}, R\&D investment \citep{Einy et al: 2015}, and bargaining \citep{Rubinstein: 1982}. Despite the complications stemming from an infinite action space, Nash equilibria are known to exist in these models. The task of equilibrium refinement theory is to identify equilibria that best shield a player against opponents making mistakes in their choice of equilibrium actions. The present paper offers a notion of perfect equilibrium that is robust against \emph{all} possible types of mistakes.
\vskip6pt

{\eightrm EARLIER CONTRIBUTIONS.} \quad
Previous generalizations of perfect equilibrium can be divided into two main categories. The first category is characterized by the use of completely mixed strategies. It models players' mistakes by probability measures that assign positive probability to each open set of an appropriately chosen Borel field.\footnote{The exception being \cite{Reny Myerson 2020}, who also allow signals to be perturbed by Dirac measures.}\label{Footnote1} Exponents of this category include weak and strong perfect equilibrium \citep{Simon Stinchcombe:1995}, perfect equilibrium \citep{Ignacio: 1995},
perfect conditional $\veps$-equilibrium \citep{Reny Myerson 2020}, sequential equilibrium \citep{Greinecker: 2025}, and behavioral perfect equilibrium \citep{Elnaz: 2016}. Another closely related, and very relevant, contribution is \cite{Marinacci: 1997}, defining perfect equilibrium for games with finitely additive strategies.
\vskip6pt

The second category takes a finitistic approach. Solutions to infinite games are constructed by taking limits of solutions of finite approximations of the original game. Exponents include limit-of-finite and anchored-limit-of-finite perfect equilibrium \citep{Simon Stinchcombe:1995}, and global-limit-of-finite perfect equilibrium \citep{Elnaz: 2013}. \cite{Harris 2000} provides an overview, with a recent contribution by \cite{Dilme 2024}. The finitistic approach is not without drawbacks, as is illustrated by for instance Example 2.1 in \cite{Reny Myerson 2020} and Example 4 in \cite{Elnaz: 2013}.\footnote{
The latter features a completely mixed Nash equilibrium which fails to be a limit of solutions for any finite approximation. In the former any finitistically feasible outcome fails to be an (even approximate) Nash equilibrium.}
\vskip6pt

{\eightrm HALLMARKS AND RESULTS.} \quad
Items [1] and [2] below discuss how the approach in this paper differs from earlier contributions. Items [3], [4], and [5] provide a brief overview of the results presented in this paper.
\vskip6pt

{\bf [1] Scope of application}. \quad
The literature on refinements that are based on completely mixed strategies typically requires games to have continuous payoff functions and action spaces that are endowed with a pre-specified Borel field that is countably generated. The definition of perfection in this paper (Def \ref{def-perf}) can be applied to more general games, allowing for example for discontinuous payoff functions and non-metrizable strategy spaces (Def \ref{def-strformgame}). Our definition of perfect equilibrium is topology free in the sense that it does not require (although it allows) the action spaces to be endowed with a topology.
\vskip6pt

{\bf [2] Carrier and nets}. \quad
Our notion of perfection mandates robustness against all possible types of mistakes.
To achieve this, it defines the carrier of a strategy to be the entire collection of sets of actions to which this strategy assigns strictly positive probability. This definition of a carrier captures more information than the usual notion of support is able to do.\footnote{
Our definition of a carrier allows, for example, to distinguish between the uniform measure on the unit interval, and the probability measure that selects the uniform measure with probability $\frac{1}{2}$, and the Dirac measure on 1 with probability $\frac{1}{2}$.} Perfect equilibrium then requires that, for each such carrier, there exists a perturbed strategy profile close to the equilibrium which has at least that carrier, and against which a best response exists that is also close to the equilibrium.\footnote{This is a slight reformulation of Definition \ref{def-perf}.
The definition itself is phrased in terms of nets of perturbed strategy profiles and corresponding best responses. See also Thm \ref{prop: characterization}.} This way, perfection shields against all types of mistakes. By using nets rather than sequences the definition also extends to non-metric spaces.
\vskip6pt

{\bf [3] Existence}. \quad
Our definition admits a perfect equilibrium for a large class of games, where it only selects Nash equilibria (Thm \ref{theorem-mainexistence}). Existence is in particular guaranteed for two well-known types of games: (1) games with compact metric action spaces, continuous payoffs, and countably additive mixed strategies (Thm \ref{theo: existence finadd}), and (2) games with arbitrary action spaces, payoff functions that are uniform limits of simple functions, and finitely additive strategies (Thm \ref{theo: existence countably additive}).
\vskip6pt

{\bf [4] Relation to earlier notions}. \quad In the case of finitely many actions, our notion of perfect equilibrium coincides with that of \cite{Selten: 1975} (Thm \ref{theo: finite games}). For games with countably additive strategies, a perfect equilibrium in our sense is a weak perfect equilibrium as defined in \cite{Simon Stinchcombe:1995} (Thm \ref{theo: weak perfect}). For games with finitely additive strategies, all perfect equilibria in the sense of \cite{Marinacci: 1997} are also perfect in our sense (Thm \ref{theo: M implies perfect}).
\vskip6pt

{\bf [5] Game theoretic desiderata}. \quad
Weakly dominant strategy equilibria are perfect in the sense of the present paper (Thm \ref{theo: dominant is perfect}). Under a mild condition, perfect equilibrium strategies are limits of undominated strategies, and any player who has a dominant strategy also uses one in every perfect equilibrium (Thm \ref{theo: limit undominated}). For strategic form games with finitely additive strategies, perfect equilibrium satisfies a weak form of the invariance requirement (Thm \ref{perfect equilibrium is invariant}). In the finitely additive framework, our results show that perfect equilibrium allows to identify payoff equivalent actions. Example \ref{example: PE not invariant} clarifies that, without further restrictions, the same is not true in the countably additive framework.\footnote{This was already observed in \cite{Elnaz: 2013}.} 
\vskip6pt

{\eightrm SETUP OF THE PAPER.} \quad
Section \ref{Section: Preliminaries} lays out the preliminaries. Section \ref{Section: Perfect equilibrium} presents the definition of perfect equilibrium. Section \ref{Section: Examples} motivates the key aspects of the definition. Section \ref{Section: Existence} and Section \ref{Section: Standard frameworks} show the existence of perfect equilibrium for two well-known classes of strategic form games. Game theoretic implications of our definition follow in the form of admissibility in Section \ref{Section: Admissibility} and of invariance in Section \ref{Section: Invariance}. Section \ref{Section: Earlier Notions} relates our notion of perfection to other ones in the literature. Section \ref{Section: Conclusion} concludes.
\vskip6pt

{\eightrm RELATED LITERATURE.} \quad
The theory on equilibrium refinement arguably started with \cite{Wu Jiang: 1962} on essential equilibrium and \cite{Harsanyi: 1973} on regular equilibrium. But these refinements do not always exist for strategic form games with finitely many actions. Early notions of equilibrium refinement that do exist for all games are perfect equilibrium \citep{Selten: 1975} and proper equilibrium \citep{Myerson: 1978}. Stronger such notions, again not existing for every game, are strictly perfect equilibrium (see for example \cite{Okada: 1981} and \cite{Jansen Jurg: 1994}) and strictly proper equilibrium \citep{Eric: 1991}. For games in extensive form, refinements of Nash equilibrium include sequential equilibrium \citep{Kreps Wilson: 1982}, perfect conditional $\veps$-equilibrium \citep{Reny Myerson 2020}, and behavioral perfect equilibrium \citep{Elnaz: 2016}.
For overviews on equilibrium refinements, see \citet{Eric: 1991,{Eric: 2002}}. The literature on strategic stability of equilibrium \citep{Kohlberg Mertens: 1986, Mertens: 1989, Mertens: 1991} takes a systematic approach to equilibrium selection based on pre-specified game theoretic desiderata. \cite{Hillas Kohlberg: 2002}, and \cite{Eric: 1991,Eric: 2002} discuss this literature in detail.
\vskip6pt

More recently, \cite{Chen: 2022} introduced robust perfect equilibrium as a refinement of Nash equilibrium in games with infinitely many players.
\cite{Weibull: 2014} apply refinement techniques in the context of continuum action spaces. For recent literature on games with finitely additive strategies, see for example \cite{Zseleva: 2018}.

\section{Preliminaries}\label{Section: Preliminaries}

This section lays out mathematical concepts used repeatedly throughout this paper. This includes nets as a generalization of sequences, and the definition of finitely additive and countably additive probability measures.
\vskip6pt

{\eightrm NETS} \quad
Let $I$ be a non-empty set, and let $\le$ be a binary relation on $I$. When $\alpha \le \beta$ and $\alpha \not= \beta$, we write $\alpha < \beta$. The pair $(I, \le)$ is a directed set when
\vspace{-.15cm}
\begin{itemize}
\item[{[1]}] (reflexivity) for every $\alpha \in I$ we have $\alpha\leq \alpha$,
\vspace{-.15cm}
\item[{[2]}] (transitivity) for every $\alpha, \beta,\gamma \in I$ such that $\alpha \le \beta$ and $\beta \le \gamma$ we have $\alpha \le \gamma$,
\vspace{-.15cm}
\item[{[3]}] (every pair has an upper bound) for every $\alpha, \beta \in I$ there is $\gamma \in I$ such that $\alpha \leq \gamma$ and $\beta \leq \gamma$, and
\vspace{-.15cm}
\item[{[4]}] (there is no maximal element) for every $\alpha\in I$ there is $\beta \in I$ with $\alpha <\beta$.
\end{itemize}
\vskip6pt
\vspace{-.15cm}
Let $X$ be a Hausdorff space. A \emph{net} on $X$, usually denoted as $x = (x_\alpha)_{\alpha \in I}$, is a map $x \colon I \rightarrow X$ from a directed set $(I, \le)$ to $X$. As an example, every real-valued sequence $(x_n)_{n \in \na}$ is a net, where $I = \na$ is the directed set with the usual order.
\vskip6pt

Let $x = (x_\alpha)_{\alpha \in I}$ be a net on $X$, and let $z \in X$. We say that $x$ converges to $z$ when for every open set $U \ni z$ there is an $\alpha \in I$ such that $\beta \ge \alpha$ implies $x_\beta \in U$.
\vskip6pt

Let $(J, \le)$ be a directed set. A net $y = (y_\beta)_{\beta \in J}$ is a subnet of
a net $x = (x_\alpha)_{\alpha \in I}$ if there exists a map $\phi \colon J \rightarrow I$ such that 
\vspace{-.15cm}
\begin{itemize}
\item[{[1]}] for every $\beta \in J$ we have $y_\beta = x_{\phi(\beta)}$, and
\vspace{-.15cm}
\item[{[2]}] for every $\alpha \in I$ there is a $\beta \in J$ such that for every $\gamma \in J$ with $\gamma\ge \beta$ we have $\phi(\gamma) \ge \alpha$.
\end{itemize}
\vspace{-.15cm}
We say that $z\in X$ is a \emph{cluster point} of the net $x = (x_\alpha)_{\alpha \in I}$ if for every open set $U \ni z$ and every $\alpha \in I$ there exists $\beta \in I$ such that $\beta\geq \alpha$ and $x_\beta \in U$. By Theorem 2.16 in \cite{Aliprantis: 2006}, $z \in X$ is a cluster point of a net $x = (x_\alpha)_{\alpha \in I}$ precisely when there is a subnet $y = (y_\beta)_{\beta \in J}$ of $x$ that converges to $z$. 
\vskip6pt

{\eightrm PROBABILITY MEASURES} \quad
Let $X$ be any set. A collection $\F$ of subsets of $X$ is called a field on $X$ if [1] $\emptyset \in \F$, [2] $X \setminus F \in \F$ for all $F \in \F$, and [3] for every pair of sets $F, G \in \F$ it holds that $F \cup G \in \F$.
\footnote{Throughout this paper we assume that $\F$ is a field on $X$ that contains each singleton set $\{ x \}$, where $x \in X$.}
A finitely additive probability measure ({\it charge}) is a function $\sigma \colon \F \rightarrow [0, 1]$ such that [1] $\sigma(\emptyset) = 0$ and $\sigma(X) = 1$, and [2] for every pair of disjoint sets $F, G \in \F$ it holds that $\sigma(F \cup G) = \sigma(F) + \sigma(G)$. We denote by $\Xi(X)$ the set of all such probability measures.\footnote{A more precise notation is $\Xi(\F)$. But, the context always clarifies which field $\F$ on $X$ is used.}
Given a point $z \in X$, the Dirac measure on $z$ is denoted by $\delta(z) \in \Xi(X)$.
\vskip6pt

A field $\F$ is called a sigma-field if $\bigcup\limits_{j \in \N} E_j \in \F$ for every countable collection $(E_j)_{j \in \N}$ of sets in $\F$. If $X$ is a topological space, the smallest sigma-field that contains all open sets is called the Borel sigma-field, denoted by $\B(X)$.
When $\F$ is a sigma-field, a probability measure $\sigma \colon \F \rightarrow [0, 1]$ is called a countably additive probability measure if
$
\sigma \Big(\bigcup_{j \in \N} E_j\Big) \,=\,\sum _{j \in \N} \sigma(E_j)
$
for every countable collection $(E_j)_{j \in \N}$ of pairwise disjoint sets in $\F$. We denote the set of all countably additive probability measures by $\Delta(X)$.
For $\mu, \nu \in \Delta(X)$, we say that $\mu$ is absolutely continuous with respect to $\nu$ if for every $F \in \F$,
$\nu(F) = 0$ implies that $\mu(F) = 0$. We say that $\mu$ and $\nu$ are mutually singular if there is $F \in \F$ with $\mu(F) = 0$ and $\nu(F) = 1$.
\vskip6pt

Several examples in this paper draw on two well-known classes of games. The following paragraphs set up the terminology and notation needed to this end.
\vskip6pt

{\eightrm FINITELY ADDITIVE MEASURES ON THE NATURAL NUMBERS} \quad
The set $\Xi(\na)$ of all charges---finitely additive probability measures---on the set $\na$ of natural numbers, endowed with the field $\F = 2^{\na}$.
The set $\Xi(\na)$ is endowed with the Tychonov topology.
\footnote{A precise definition of the Tychonov topology, also in more generality for any arbitrary set of finitely additive probability measures, is provided in Section \ref{Section: Standard frameworks}.}
Note that the resulting topological space $\Xi(\na)$ is compact, but not metrizable.
\vskip6pt

A charge $\kappa \in \Xi(\na)$ is called diffuse when $\kappa(\{k\}) = 0$ for all $k \in \na$.
For every $\kappa \in \Xi(\na)$ there is a unique way to write $\kappa$ as a convex combination of a diffuse charge and a countably additive probability measure on $\na$. (See \cite{Rao Rao 1983} for reference.)
\vskip6pt

{\eightrm COUNTABLY ADDITIVE MEASURES ON THE UNIT INTERVAL} \quad
The set $\Delta[0, 1]$ of all countably additive probability measures on the unit interval $[0, 1]$, endowed with the Borel sigma-field $\B[0, 1]$.
The set $\Delta[0, 1]$ is endowed with the topology of weak convergence, metrized by the Prokhorov metric.\footnote{Precise definitions can be found in Section \ref{Section: Standard frameworks}.} Throughout this paper, we consistently use $\lambda$ to indicate the uniform measure on the unit interval, i.e., the Lebesgue measure defined on the Borel sigma-field $\B[0, 1]$.

\section{Perfect equilibrium}
\label{Section: Perfect equilibrium}

This section defines our notion of perfect equilibrium. It starts with a precise description of the class of games to which this definition applies. Next, it presents our definition of perfect equilibrium within this class of games, followed by a characterization of perfect strategy profiles. A brief discussion relates our notion of perfect equilibrium to earlier such notions in the literature.\footnote{A more detailed discussion of the relations is deferred to Section \ref{Section: Earlier Notions}.}

\vspace{-.25cm}
\subsection{Strategic form games} \label{PE}

The next definition of a strategic form game provides the bare minimum we need in order to be able to apply our definition of perfect equilibrium.

\begin{definition}[\textbf{strategic form game}]\label{def-strformgame}
A \emph{strategic form game} is a tuple $\Gamma = (N, A, \F, \Sigma, U)$, where
\begin{itemize}
\vspace{-.15cm}
\item $N = \{ 1, \ldots, n \}$ is the set of players. 
\vspace{-.15cm}
\item $A = \prod_{i\in N} A_i$, where $A_i$ is a non-empty set of actions for each player $i \in N$. 
\vspace{-.15cm}
\item $\F = (\F_1, \ldots, \F_n)$, where $\F_i$ is a field on $A_i$ containing each singleton $\{a_i\}$, where $a_i\in A_i$, for each player $i \in N$.
\vspace{-.15cm}
\item $\Sigma = \prod_{i\in N} \Sigma_i$, where $\Sigma_i\subseteq \Xi(A_i)$ is a non-empty set of available strategies for each player $i \in N$, endowed with a (Hausdorff) topology.
\vspace{-.15cm}
\item $U = (U_1, \ldots, U_n)$, where $U_i \colon \Sigma \rightarrow \re$ is a payoff function for each player $i \in N$.
\vspace{-.15cm}
\end{itemize}
\end{definition}
\vspace{-.35cm}

We use the following terminology.
A \emph{finitely additive strategy} for player $i\in N$ is a finitely additive probability measure $\sigma_i \in \Xi(A_i)$. Let $\Xi_i = \Xi(A_i)$ denote the set of all finitely additive strategies of player $i$. Let $\Xi=\prod_{i\in N} \Xi_i$ denote the set of all finitely additive strategy profiles.
When $\F_i$ is a sigma-field, a countably additive probability measure $\sigma_i \in \Xi_i$ is called a \emph{countably additive strategy}. Let $\Delta_i = \Delta(A_i)$ denote the set of countably additive strategies of player $i$, and let $\Delta=\prod_{i\in N} \Delta_i$ denote the set of countably additive strategy profiles. In some results and most examples in this paper, the set $\Sigma$ of available strategy profiles is either $\Sigma=\Xi$ or $\Sigma=\Delta$.
\vskip6pt

{\bf Remark.} \quad
Note that we do not require the payoff function $U_i$ of a player $i\in N$ to be the expected value of a payoff function $u_i \colon A \rightarrow \mathbb{R}$ defined on the set of action profiles. In most of the examples we consider in the paper, $U_i$ is the expected value of such a function $u_i$. But the definition of perfect equilibrium does not rely on it, and indeed in footnote \ref{footnote Ui} the payoff function $U_i$ is defined directly on strategy profiles.

\vspace{-.25cm}
\subsection{A definition of perfect equilibrium}\label{subs_def-pe}

This section defines perfect equilibrium for strategic form games. Section \ref{sect: relations} relates this new definition to earlier such notions in the literature, and Section \ref{Section: Examples} elaborates on some of its key aspects. Let $\Gamma = (N, A, \F, \Sigma, U)$ be a strategic form game as in Definition \ref{def-strformgame}. We start by defining best responses and Nash equilibrium.
\vskip6pt

\textbf{Best response.} For a strategy profile $\sigma \in \Sigma$ and a strategy $\tau_i \in \Sigma_i$ for some player $i\in N$, by $(\tau_i  \mid \sigma)$ we denote the strategy profile where every player $j \not= i$ adheres to strategy $\sigma_j$, while player $i$ uses $\tau_i$. A strategy $\tau_i$ is called a best response to $\sigma$ if $U_i(\tau_i \mid \sigma ) \ge U_i(\tau'_i \mid \sigma)$ for every strategy $\tau'_i \in \Sigma_i$. A strategy profile $\tau\in \Sigma$ is called a best response to a strategy profile $\sigma\in \Sigma$ if $\tau_i$ is a best response to $\sigma$ for each player $i\in N$. \vskip6pt

\begin{definition}[\textbf{Nash equilibrium}]\label{def-NE}
A strategy profile $\sigma\in\Sigma$ is a \emph{Nash equilibrium} if $\sigma$ is a best response to $\sigma$. 
\end{definition}

We are now ready to define perfection of a Nash equilibrium. The definition is based on a new interpretation of what the carrier of a strategy is, leading to the notion of a completely mixed net. We first formally define these notions. For a more elaborate motivation, see Section \ref{Section: Examples}.
\vskip6pt

\textbf{Carrier.} 
$
\C_i(\sigma_i) = \{ E_i \in \F_i \mid \sigma_i(E_i) > 0 \}
$ defines the carrier of a strategy $\sigma_i \in \Sigma_i$. A collection $\C_i \subseteq \F_i$ is called a carrier for player $i$ if there is a strategy $\sigma_i \in \Sigma_i$ with $\C_i = \C_i(\sigma_i)$. We write $\C(\sigma) = \prod_{i\in N} \C_i(\sigma_i)$ and $\C = \prod_{i\in N} \C_i$. We say $\C$ is a carrier if $\C_i$ is a carrier for each player $i$. If $\C_i(\sigma_i) = \F_i \setminus \{ \emptyset \}$, we say that $\sigma_i$ has maximal carrier.
\vskip6pt

Note that in general there does not exist a strategy with maximal carrier (see also Section \ref{subsection: net motivation}). Therefore, we use the notion of a completely mixed net in order to model robustness of perfect equilibrium against any possible type of mistakes.
\vskip6pt

\textbf{Completely mixed net.} A net $(\sigma_\alpha)_{\alpha \in I}$ on the set $\Sigma$ of available strategy profiles is called \emph{completely mixed} if for every carrier $\C$ there is $\alpha \in I$ such that $\C \subseteq \C(\sigma_\beta)$ for every $\beta \ge \alpha$.
\vskip6pt

Thus, a net of available strategy profiles is completely mixed if eventually  every possible carrier is included, when we advance sufficiently far into the net. This notion of a completely mixed net is a crucial ingredient in the following definition.

\begin{definition}[\textbf{perfect strategy profile}]\label{def-perfstr}
A strategy profile $\sigma\in\Sigma$ is called \emph{perfect} if there is a completely mixed net $(\sigma_\alpha)_{\alpha \in I}$ on $\Sigma$ and a net $(\kappa_\alpha)_{\alpha \in I}$ on $\Sigma$ such that
\vspace{-.15cm}
\begin{itemize}
\item[{\rm [1]}] $(\sigma_\alpha)_{\alpha \in I}$ converges to $\sigma$,
\vspace{-.15cm}
\item[{\rm [2]}] $(\kappa_\alpha)_{\alpha \in I}$ converges to $\sigma$, and
\vspace{-.15cm}
\item[{\rm [3]}] for each $\alpha \in I$, $\kappa_\alpha$ is a best response to $\sigma_\alpha$.
\end{itemize}
\end{definition}

The definition of a perfect strategy profile generalizes the best-response type of definition of perfect equilibrium for strategic form games with finite action spaces.

\begin{definition}[\textbf{perfect equilibrium}]\label{def-perf} 
A strategy profile $\sigma\in\Sigma$ is called a \emph{perfect equilibrium} if $\sigma$ is a Nash equilibrium and $\sigma$ is perfect.
\end{definition}

\vspace{-.25cm}
\subsection{A characterization of perfect strategy profiles}

Perfect strategy profiles are a prerequisite of perfect equilibrium. This makes it desirable to attain a characterization of perfect strategy profiles. What sets our characterization apart is that it specifies conditions for each possible carrier separately, rather than relying on conditions along (completely mixed) nets of carriers as in Definition \ref{def-perfstr}. The characterization given here is applied throughout the paper to verify perfection of strategy profiles, especially in examples.

\begin{theorem} \label{prop: characterization}
Let $\sigma\in\Sigma$ be a strategy profile. Equivalent are:
\begin{itemize}
\vspace{-.15cm}
\item[{\rm [1]}] The strategy profile $\sigma$ is perfect.
\vspace{-.15cm}
\item[{\rm [2]}] For every carrier $\C$ and every open set $U \ni \sigma$
there are $\tau, \kappa \in U$ such that $\C \subseteq \C(\tau)$ and $\kappa$ is a best response to $\tau$.
\vspace{-.15cm}
\item[{\rm [3]}] For every carrier $\C$  there are nets $(\sigma_\alpha)_{\alpha \in J}$ and $(\kappa_\alpha)_{\alpha \in J}$ converging to $\sigma$ such that for each $\alpha \in J$ it holds that $\C \subseteq \C(\sigma_\alpha)$ and $\kappa_\alpha$ is a best response to $\sigma_\alpha$.
\end{itemize}
\end{theorem}

\noindent
\textbf{Proof of [1] $\Rightarrow$ [3]:} Let the nets $(\sigma_\alpha)_{\alpha \in I}$ and $(\kappa_\alpha)_{\alpha \in I}$ be as in Definition \ref{def-perfstr} of perfect strategy profile. Take any carrier $\C$. Since $(\sigma_\alpha)_{\alpha \in I}$ is completely mixed, there is $\beta \in I$ with $\C \subseteq \C(\sigma_\alpha)$ for every $\alpha \ge \beta$. Take $
J = \{ \alpha \in I \mid \alpha \ge \beta \}.
$
Then $J$ is a directed set, and the nets $(\sigma_\alpha)_{\alpha \in J}$ and $(\kappa_\alpha)_{\alpha \in J}$ satisfy the requirements in [3].
\vskip6pt

\noindent
\textbf{Proof of [3] $\Rightarrow$ [2]:} Take any carrier $\C$. Take nets $(\sigma_\alpha)_{\alpha \in J}$ and $(\kappa_\alpha)_{\alpha \in J}$ that satisfy the requirements in [3]. Take any open $U \ni \sigma$.
Then there is $\beta \in J$ such that $\sigma_\beta \in U$ and $\kappa_\beta \in U$. Then
$\tau = \sigma_\beta$ and $\kappa = \kappa_\beta$ satisfy the requirements in [2].
\vskip6pt

\noindent
\textbf{Proof of [2] $\Rightarrow$ [1]:} Take 
$
I = \{ (\C, U) \mid \C \hbox{ is a carrier and } U \ni \sigma \hbox{ is open} \}
$
ordered by $(\C, U) \le (\D, V)$ precisely when both $\C \subseteq \D$ and $U \supseteq V$.
Then $I$ is a directed set. For every $\alpha = (\C, U) \in I$ we take $\tau, \kappa \in U$ as in [2], and define
$\sigma_\alpha = \tau$ and $\kappa_\alpha = \kappa$. This defines nets $(\sigma_\alpha)_{\alpha \in I}$ and $(\kappa_\alpha)_{\alpha \in I}$ with the properties required by [1].
\hfill \qed

\vspace{-.25cm}
\subsection{Relation to existing notions of perfect equilibrium} \label{sect: relations}

We now state how our notion of perfect equilibrium relates to the notion of perfect equilibrium in \cite{Selten: 1975}, the notion of weak perfect equilibrium in \cite{Simon Stinchcombe:1995}, and the notion of perfect equilibrium in \cite{Marinacci: 1997}. Section \ref{Section: Earlier Notions} analyzes these relations in more detail, involving proofs and definitions.
\vskip6pt

For a game in strategic form with finite action sets, the set of perfect equilibria in the sense of \cite{Selten: 1975} coincides with the the set of perfect equilibria in our sense (cf. Thm \ref{theo: finite games}).
\vskip6pt

For a game in strategic form with compact metric action spaces and continuous payoff functions, our definition yields a nonempty and compact set of perfect equilibria. All these equilibria are also weak perfect equilibria in the sense of \cite{Simon Stinchcombe:1995} (cf. Thm \ref{theo: weak perfect}). When there are only two players, and one player has only two actions, the set of perfect equilibria in our sense (cf. Thm \ref{theo: two players}) coincides with the set of weak perfect equilibria. It remains an open question whether this is true in general.
\vskip6pt

For a game in strategic form with finitely additive strategies, any equilibrium that is perfect in the sense of \cite{Marinacci: 1997} is also a perfect equilibrium in our sense (cf. Thm \ref{theo: M implies perfect}).\footnote{In \cite{Marinacci: 1997} perfection of Nash equilibrium is a strong requirement. In our setting, when all singleton action sets are measurable, perfect equilibrium in the sense of \cite{Marinacci: 1997} only exists when all action spaces are at most countable.}
Theorem \ref{theo: special Marinacci 1} provides a partial result on the reverse implication.

\section{Motivation and illustrative examples}
\label{Section: Examples}

This section motivates our definition of perfect equilibrium. It hones in on certain key aspects, such as that of a carrier, the usage of nets rather than sequences, and the applicability to general games. Examples illustrate the working of these aspects.

\vspace{-.25cm}
\subsection{The notion of carrier, a motivation}
\label{carrier motivation}

Our definition of carrier deviates from the usual notions of carrier or support that are used to measure to which extent a strategy mixes between different actions. The following examples illustrate what mandates this departure to a more detailed description of carrier.
\vskip6pt

Given $A_i = [0, 1]$, the classic definition of support (i.e., the smallest compact set that has probability one) does not discriminate between the support of the uniform measure on $A_i$, and the support of the strategy that with probability half plays according to the uniform measure, and with probability half plays a Dirac measure on a singleton $x \in A_i$. Since both have support $A_i$ in the classic definition, mixing in a Dirac measure into the uniform measure does not affect the support under this definition. This stands in stark contrast to the fact that accidentally mixing in strategies like a Dirac measure\footnote{Or any type of measure that is mutually singular with the strategy one intends to play.} may well and often will have consequences for expected payoffs. Adapting the notion of carrier allows to address such consequences via differences in the respective carriers. Under our notion of a carrier, the singleton set $\{ x \}$ is not an element of the carrier of the uniform distribution, but it is an element of the carrier of the mixture of the uniform distribution and the Dirac measure on $x$.\footnote{Our approach results in an admittedly extreme form of carrier, taking into account all sets in $\F_i$ which have strictly positive probability. Although more economical definitions may well be possible, here we adopt the most extreme form possible as our working definition of carrier.}
\vskip6pt

The next examples illustrate the working of our carrier. Example A shows how our definition accommodates possible deviations to Dirac measures. Example B demonstrates that accommodating \emph{any} type of measure as a possible deviation is essential.

\vspace{-.25cm}
\subsubsection*{Example A. \quad A discontinuous coordination game}

Consider the game where the player set is $N=\{1,2\}$, the action sets are $A_1 = A_2 =[0,1]$, the strategy spaces are $\Sigma_i= \Delta(A_i)$, and the sigma-fields are $\F_i = \B[0, 1]$ for $i = 1, 2$. As specified in Section \ref{Section: Preliminaries}, each $\Sigma_i$ is endowed with the weak topology induced by the Prokhorov metric. The payoff functions are defined by
$$
u_1(a_1,a_2) = u_2(a_1,a_2) =
\begin{cases}
1 & \hbox{if } (a_1, a_2) = (1, 1) \cr
0 & \hbox{otherwise.}
\end{cases}
$$
{\bf Claim A1.}
{\it
A strategy pair $\sigma = (\sigma_1, \sigma_2) \in \Sigma$ is a perfect equilibrium of this game precisely when $\sigma_1 = \sigma_2 = \delta(1)$.}
\vskip6pt

Proof of claim A1: It is straightforward to show that $(\delta(1), \delta(1))$ is a perfect equilibrium.
Conversely, assume that some $\sigma \in \Sigma$ is a perfect equilibrium.
Then by Definition \ref{def-perfstr} there is a completely mixed net $(\sigma_\alpha)_{\alpha \in I}$ on $\Sigma$ converging to $\sigma$,
and a net $(\kappa_\alpha)_{\alpha \in I}$ on $\Sigma$ converging to $\sigma$ such that $\kappa_\alpha$ is a best response to $\sigma_\alpha$ for each $\alpha \in I$.
\vskip6pt

Note that there is a carrier $\C$ with $\{1\} \times \{1\} \in \C$. So, since $(\sigma_\alpha)_{\alpha \in I}$ is completely mixed, there is $\gamma \in I$ such that  $\{1\} \times \{1\} \in \mathcal{C}(\sigma_\beta)$ for every $\beta \geq \gamma$. So, for every $\beta \geq \gamma$, $\kappa_\beta$ is a best response to $\sigma_\beta$ whose carrier includes $\{1\} \times \{1\}$, and hence $\kappa_\beta = (\delta(1), \delta(1))$.
So $(\kappa_\alpha)_{\alpha \in I}$ converges to $(\delta(1), \delta(1))$. Hence, $\sigma = (\delta(1), \delta(1))$.
\hfill \qed
\vskip6pt

{\bf Remarks.} \quad
Let $\mu$ be any probability measure that is absolutely continuous with respect to the uniform measure $\lambda$ over $\B[0, 1]$. Then $(\mu, \mu)$ is a strong (and hence also weak) perfect equilibrium by \cite{Simon Stinchcombe:1995}. Any such equilibrium yields payoff zero for both players, illustrating the difference between Definition \ref{def-perf} and the definition of weak perfect equilibrium in \cite{Simon Stinchcombe:1995}.\footnote{
Of course, the setup of \cite{Simon Stinchcombe:1995} is designed for continuous payoff functions, so this example is not covered by their framework.}
A similar analysis can be held for any $u_1 = u_2 = \id_{Q \times Q}$, with $Q \in \F_i$ and $\lambda(Q) = 0$. This concludes Example A.
\hfill $\triangleleft$

\vspace{-.15cm}
\subsubsection*{Example B. \quad A Cantor coordination game}

The previous Example A suggests that, at least in the context of the unit interval, a weaker definition of perfect equilibrium could suffice, specifically one that only considers Dirac measures and measures that are absolutely continuous with respect to the Lebesgue measure. The next example demonstrates that this is not the case.
\vskip6pt

The game has two players. The action sets are $A_1 = A_2 = [0, 1]$ and the sigma-fields are $\F_1 = \F_2 = \B[0, 1]$. 
Let ${\mathbb D}$ denote the Cantor set. 
Define the payoff functions by
$$
u_1(a_1, a_2) = u_2(a_1,a_2) = 
\begin{cases}
1 & \hbox{if } a_1 \in {\mathbb D} \hbox{ and } a_2 \in {\mathbb D} \cr
0 & \hbox{otherwise.}
\end{cases}
$$
For each player $i = 1,2$, the strategy space $\Sigma_i$ is the collection of those strategies $\sigma_i \in \Delta[0, 1]$ for which the cumulative distribution function (cdf) of $\sigma_i$ is continuous. Note that, for $(\sigma_1, \sigma_2) \in \Sigma$, we have that
$U_1(\sigma_1, \sigma_2) = U_2(\sigma_1, \sigma_2) = \sigma_1({\mathbb D}) \cdot \sigma_2({\mathbb D})$.
Further, note that there is a strategy $\sigma_i \in \Sigma_i$ with $\sigma_i({\mathbb D}) = 1$. For example, the strategy for which its cdf is the Devil's Staircase is such a strategy.
\vskip6pt

{\bf Claim B.}
{\it
A strategy pair $(\sigma_1, \sigma_2)$ is a perfect equilibrium precisely when $\sigma_1({\mathbb D}) = \sigma_2({\mathbb D}) = 1$. In any perfect equilibrium, both players receive payoff 1.}
\vskip6pt

Proof of claim: It is clear that $\sigma_i$ is a dominant strategy precisely when $\sigma_i({\mathbb D}) = 1$.\footnote{A strategy is dominant if it is a best response against any strategy profile. See also Section \ref{Section: Admissibility}.}
Hence, by Theorem \ref{theo: dominant is perfect}, if $\sigma_1({\mathbb D}) = \sigma_2({\mathbb D}) = 1$, then $(\sigma_1, \sigma_2)$ is a perfect equilibrium.
\vskip6pt

Conversely, let $(\sigma_1, \sigma_2)$ be a perfect equilibrium. Let $(\sigma_\alpha)_{\alpha \in I}$ and $(\kappa_\alpha)_{\alpha \in I}$ be as in Definition \ref{def-perfstr}. Since there is a carrier $\C$ with ${\mathbb D} \times {\mathbb D} \in \C$, we can assume that $\sigma_{\alpha, i}({\mathbb D}) > 0$ for all $\alpha \in I$ and $i = 1, 2$. Since $\kappa_\alpha$ is a best response to $\sigma_\alpha$, it follows that $\kappa_{\alpha, i}({\mathbb D}) = 1$ for all $\alpha \in I$ and $i = 1, 2$. Hence, also $\sigma_i({\mathbb D}) = 1$ for $i = 1, 2$.
\hfill \qed
\vskip6pt

{\bf Remarks.} \quad
This game has many Nash equilibria that are not perfect. To see this, suppose that one of the players uses a strategy $\sigma_i$ with $\sigma_i({\mathbb D}) = 0$---for example $\sigma_i = \lambda$. Then both players receive payoff zero. Thus, both players using such strategies results in a Nash equilibrium. In the above game, the Cantor set  ${\mathbb D}$ can be replaced by an arbitrary uncountable compact subset $K \subseteq [0, 1]$ with empty interior.\footnote{
\label{footnote Ui}
A version of the example that does allow for Dirac measures as strategies (and hence also as deviations) can be constructed by taking $\Sigma_i = \Delta[0, 1]$, and defining 
$
U_1(\sigma_1, \sigma_2) = U_2(\sigma_1, \sigma_2) = \sigma_1^\ast({\mathbb D}) \cdot \sigma_2^\ast({\mathbb D}),
$
where $\sigma_i^\ast$ denotes the absolutely continuous part of $\sigma_i$ with respect to the uniform measure on $\mathbb D$.
Then, also $\left(\delta\left(\frac{1}{4}\right), \delta\left(\frac{1}{2}\right)\right)$ is a Nash equilibrium. But, it is not perfect in our sense.}
\hfill $\triangleleft$

\vspace{-.25cm}
\subsection{Net convergence, a concise motivation}
\label{subsection: net motivation}

Observe that with our new interpretation of a carrier, it is not always possible to have a maximal element in the collection of possible carriers, not even in the case where $A_i = [0, 1]$.\footnote{Indeed, while each singleton can act as an element of a carrier, there does not exist a charge in $\Xi(A_i)$ for which each singleton is an element of its carrier.} Therefore, ``completely mixed'' cannot be viewed as an attribute of individual strategy profiles. The next example demonstrates that sequences of ``increasingly mixed'' strategy profiles are also insufficient.

\begin{example} Sequences are not sufficient. \rm
Take $A_i = \omega_1$, the first uncountable ordinal. Consider a sequence of countably additive strategies. Because each countably additive strategy can only assign strictly positive weight to countably many elements of $\omega_1$, along the sequence, only countably many elements of $\omega_1$ are played with strictly positive probability. Since $\omega_1$ contains uncountably many elements, no such sequence can place strictly positive probability on each singleton in $\omega_1$.
\hfill $\triangleleft$
\end{example}

The previous example illustrates that, in general, sequences of strategy profiles are not sufficient to capture all possible types of mistakes. The more appropriate concept here is that of a net.\footnote{We do not claim any originality here. Nets have been used to define Nash equilibrium and refinements thereof in, among others, \cite{Stinchcombe: 2005} and \cite{Reny Myerson 2020}.}
Thus, we consider ``completely mixed'' to be a property of nets. A net is considered to be completely mixed if, when we advance sufficiently far into the net, every possible carrier eventually gets realized.

\vspace{-.25cm}
\subsection{\normalsize Application to games played with finitely additive strategies}

The definition applies to a wide range of settings. The next example illustrates this for the context of games played with finitely additive strategies, which places the analysis outside the context of metric spaces.

\begin{example} 
\label{example: PE is not NE}
Not every Nash equilibrium is perfect. \rm
There are two players. The action spaces are $A_1 = \{ T, B \}$ and $A_2 = \{ 1, 2, 3, 4, \ldots \}$, where all natural numbers are isolated points. The game is zero-sum, and the payoffs for player 1 are given by
$$
u_1 =
\left[
\begin{matrix}
\hfill 1 & \hfill \frac{1}{2} & \hfill \frac{1}{3} & \hfill \frac{1}{4} & \ldots\ldots \ \ \ \cr \cr
-1 & - \frac{1}{2} & - \frac{1}{3} & - \frac{1}{4} & \ldots\ldots \ \ \ 
\end{matrix}
\right].
$$
Observe that the game does not have Nash equilibria in countably additive strategies. We argue that, when we allow players to use finitely additive strategies, so that $\Sigma_i = \Xi_i$, Nash equilibria exist in this example.\footnote{Note that this game does have a uniquely defined payoff for finitely additive strategies. We discuss this in detail and in more generality in Section \ref{Sect: finaddstrats}.}
We also compute the set of perfect equilibria.\footnote{As is specified in Section \ref{Section: Preliminaries}, we endow $\Xi_i$ with the Tychonov topology. We discuss this in more detail in Section \ref{Section: Standard frameworks}.} It turns out that not all Nash equilibria are perfect.
\vskip6pt

{\bf Claim 1.} \quad
{\it A strategy pair $(\kappa_1, \kappa_2)$, with $\kappa_1 = (p, 1 - p)$, is a Nash equilibrium precisely when $p \ge \frac{1}{2}$ and $\kappa_2$ is a diffuse charge.}
\vskip6pt

Proof of Claim 1. \quad
The best response for player 1 is to play $T$ whenever $\kappa_2(n) > 0$ for some $n \in \na$. When $\kappa_2$ is diffuse, player 1 gets payoff zero for any strategy he chooses. So, in this case player 1 is indifferent between any of his strategies. The best response for player 2 against $(p, 1-p)$ is to play action $1$ when $p < \frac{1}{2}$, and to play any diffuse charge when $p > \frac{1}{2}$. When $p = \frac{1}{2}$, player 2 is indifferent.
\hfill \qed
\vskip6pt

{\bf Claim 2.} \quad
{\it A strategy pair $(\kappa_1, \kappa_2)$, with $\kappa_1 = (p, 1 - p)$, is a perfect equilibrium precisely when $p = 1$ and $\kappa_2$ is a diffuse charge.}
\vskip6pt

Proof of Claim 2. \quad
By Claim 1, we know that in any (perfect) equilibrium $\kappa_2$ is a diffuse charge. Further, since $T$ strictly dominates $B$ for each action of player 2, we have $p = 1$ in any perfect equilibrium.
\vskip6pt

We argue that such a pair of strategies is indeed a perfect equilibrium. This follows from the next two observations. First, against any mixed strategy $(p, 1-p)$ of player 1 with $p > \frac{1}{2}$, the set of best responses of player 2 is the set of diffuse charges. Second, against any mixed strategy of player 2, $p = 1$ is a best response.
\hfill $\triangleleft$
\end{example}

\section{Existence}\label{Section: Existence}

This section discusses the existence of perfect equilibrium within the framework set up in Section \ref{Section: Perfect equilibrium}. Theorem \ref{theorem-mainexistence} is the main result in this section, which provides general conditions under which a game admits a perfect equilibrium as in Definition \ref{def-strformgame}.

\begin{theorem} \label{theo: perfect eq}
Let $\Gamma = (N, A, \F, \Sigma, U)$ be a strategic form game as in Definition \ref{def-strformgame}. Suppose that the payoff function $U_i$ of each player $i\in N$ is continuous. Then, every perfect strategy profile is a Nash equilibrium.
\end{theorem}

\noindent
Proof. \quad
Let $\sigma\in \Sigma$ be a perfect strategy profile. Fix a player $i\in N$ and a strategy $\tau_i \in \Sigma_i$. We need to show that $U_i(\sigma_i \mid \sigma) \ge U_i(\tau_i \mid \sigma)$.
\vskip6pt

\noindent
Since $\sigma$ is a perfect strategy profile, there are nets $(\sigma_\alpha)_{\alpha \in I}$ and $(\kappa_\alpha)_{\alpha \in I}$ as specified in Definition \ref{def-perfstr}. As such, $\kappa_{\alpha, i}$ is a best response to $\sigma_\alpha$, and hence $U_i(\kappa_{\alpha,i} \mid \sigma_\alpha) \ge U_i(\tau_i \mid \sigma_\alpha)$.
Since $U_i$ is continuous, and both nets $(\sigma_\alpha)_{\alpha \in I}$ and $(\kappa_\alpha)_{\alpha \in I}$ converge to $\sigma$, by taking limits we have  $U_i(\sigma_i \mid \sigma) \ge U_i(\tau_i \mid \sigma)$.
\hfill \qed
\vskip6pt

\begin{theorem}
[\textbf{Existence of perfect equilibrium}]
\label{theorem-mainexistence}
Let $\Gamma$ be a strategic form game as in Definition \ref{def-strformgame}. Suppose that $\Sigma$ is a compact and convex subset of a locally convex topological vector space, and that $U_i$ is multi-linear and continuous for each player $i\in N$. Then the set of perfect equilibria of $\Gamma$ is non-empty and compact.
\end{theorem}

\noindent
Outline of the proof. \quad
The proof is divided in three steps. Step A shows that a game with perturbed strategy spaces, as defined in \cite{Kohlberg Mertens: 1986} for finite games in strategic form, admits a Nash equilibrium.\footnote{The use of perturbed strategy spaces to prove existence of perfect equilibrium dates back to \cite{Selten: 1975}.} 
This step is a direct application of the  Kakutani-Fan-Glicksberg fixed-point Theorem (cf.\ Thm \ref{theo: Glicksberg}).
Step B takes a fixed carrier $\C$, and writes PS$(\C)$ for the set of limit points of all nets $(\sigma_\alpha)_{\alpha \in I}$ and $(\kappa_\alpha)_{\alpha \in I}$ as in [3] of Theorem \ref{prop: characterization}. Thus, PS$(\C)$ is compact. We use the result of Step A to show that PS$(\C)$ is non-empty.
In Step C we consider the intersection 
\begin{equation*}
\text{PS}\,:=\,\bigcap_{\C\text{ is a carrier}}\ \text{PS}(\C)    
\end{equation*} 
According to Theorem \ref{prop: characterization}, $\hbox{PS}$ is exactly the set of perfect strategy profiles. We use Step B to show that $\hbox{PS}$ is compact and non-empty. In this step we rely on a specific characterization of compact spaces related to the finite intersection property.
\hfill \qed
\vskip6pt

For brevity and clarity of exposition, the proof of Theorem \ref{theorem-mainexistence} is based on Theorem \ref{prop: characterization}, item [3]. Building a proof directly from Definition \ref{def-perfstr} is also possible. In such a proof, rather than constructing sequences for a given carrier $\C$, nets have to be constructed on the directed set $I$ of pairs $(\C, \veps)$, where $\C$ is a carrier and $\veps > 0$ is the extent to which a perturbation is mixed into the intended choice of strategy of a player. 


\section{Two classic classes of games}\label{Section: Standard frameworks}

This section presents two well-known classes of games to which our Definition \ref{def-perf} of perfect equilibrium applies. In these two classes many of our results are valid, such as Theorem \ref{theorem-mainexistence} regarding the existence of perfect equilibrium, the results on admissibility in Section \ref{Section: Admissibility}, and the results on invariance in Section \ref{Section: Invariance}.
\vskip6pt

In the first class of games, the set $\Sigma$ of strategy profiles is the set $\Xi$ of all finitely additive strategy profiles. The topology on $\Sigma = \Xi$ is the Tychonov topology. This class of games is used in \cite{Marinacci: 1997}. The second class of games is identical to the one in \cite{Simon Stinchcombe:1995}, where the set of strategy profiles $\Sigma$ is the set $\Delta$ of all countably additive strategy profiles, endowed with the Prohorov topology.
\vskip6pt

We discuss each of these two choices, $\Sigma = \Xi$ and $\Sigma = \Delta$, together with the associated respective topologies. For each choice we also give sufficient conditions to guarantee existence of perfect equilibrium. We present several examples to illustrate our results.

\vspace{-.25cm}
\subsection{Expected payoffs}\label{subs_exp}

In both classic settings each payoff function $U_i$ on strategy profiles is the expectation of a payoff function $u_i$ on the action profiles. Consider a game $\Gamma^\ast = (N, A, \F, \Sigma, u)$, where $N$, $A$, $\F$, and $\Sigma$ are as in Definition \ref{def-strformgame}, and $u = (u_1, \ldots, u_n)$, where $u_i \colon A \rightarrow \re$ is a bounded payoff function for each player $i\in N$. 
\vskip6pt

\begin{definition}\label{def-viable}
A bounded payoff function $u_i \colon A\to\re$ is \emph{viable} if $u_i$ is integrable with respect to each strategy profile $\sigma \in \Sigma$.
\end{definition}

When $u_i$ is viable, the expected payoff function is denoted by $U_i \colon \Sigma \rightarrow \re$. It is well-known that, if $u_i$ is viable, then $U_i$ is multi-linear. A short proof of this fact is provided in Lemma \ref{viable is multilinear} in Appendix A of the Supplemental Appendix.
\vskip6pt

A strategic form game $\Gamma = (N, A, \F, \Sigma, U)$ as in Definition \ref{def-strformgame} is called the \emph{mixed extension} of a game $\Gamma^\ast = (N, A, \F, u)$, if each $U_i$ is the expected payoff function of the payoff function $u_i \colon A \rightarrow \re$.
If for a game $\Gamma^\ast = (N, A, \F, u)$, for each payoff function $u_i$ and each strategy profile $\sigma \in \Sigma$ the expected payoff $U_i(\sigma)$ exists, we identify $\Gamma^\ast = (N, A, \F, u)$ with its associated mixed extension $\Gamma = (N, A, \F, \Sigma, U)$. We emphasize this identification by writing $\Gamma = (N, A, \F, \Sigma, u)$. This slight abuse of notation thus indicates that for each payoff function $u_i$, the expected payoff $U_i(\sigma)$ exists for each strategy profile $\sigma \in \Sigma$.

\vspace{-.25cm}
\subsection{Games played with finitely additive strategies} \label{Sect: finaddstrats}

One of the two showcase examples to which our theoretical framework applies is the collection of all finitely additive probability measures, endowed with Tychonov topology. This coincides with the setting in \cite{Marinacci: 1997}. 
\vskip6pt

A set $R \subseteq A$ is called a rectangle if there are sets $R_i \in \F_i$ for each $i \in N$ such that
$R = \prod_{i\in N} R_i$. A function $f_i \colon A \rightarrow \re$ is called \emph{simple} if there are rectangles
$R^1, \ldots, R^k \subseteq A$ and real numbers $c_1, \ldots, c_k$ such that
$
f_i = \sum_{\ell=1}^k c_\ell \cdot \id_{R^\ell}.
$

\begin{definition}\label{def-strformgamefinad} \rm
A strategic form game $\Gamma = (N, A, \F, \Xi, u)$ is called a \emph{strategic form game with finitely additive strategies} if the following two conditions hold.\footnote{Note that the action sets $A_i$ are arbitrary, without any further initial structure defined on them.}

\begin{itemize}
\item[{[1]}] Each $u_i$ is bounded, and a uniform limit of simple functions: for every $\veps > 0$ there is a simple function $f_i$ such that
$
\left| u_i(a) - f_i(a) \right| < \veps  \hbox{, for all } a \in A.
$
\item[{[2]}] The topology on $\Xi$ is the Tychonov topology: the smallest topology that contains all basis sets.
A set $V \subseteq \Xi$ is a basis set if there are $\kappa \in \Xi$, $\veps > 0$, and rectangles $R^1, \ldots, R^\ell$ such that
$$
V = \left\{ \tau \in \Xi \ \mid \ \left| \kappa(R^k) - \tau(R^k) \right| < \veps 
\quad \hbox{ for all } k = 1, \ldots, \ell \right\}.
$$
\end{itemize}
\end{definition}

{\bf Remarks.} \quad
As observed earlier, when $A$ is infinite, the Tychonov topology on $\Xi$ is not metrizable. For a game with finitely additive strategies, each $U_i$ exists, and each $U_i$ is multi-linear and continuous in the Tychonov topology. For comparison, we refer to (the proof of) Theorem 1 in \cite{Marinacci: 1997}. Detailed proofs are provided in Lemma \ref{theo: existence for finite additivity} in Appendix A of the Supplemental Appendix.
\hfill $\triangleleft$
\vskip6pt

Theorem 1 in \cite{Marinacci: 1997} shows that the set of Nash equilibria is non-empty and compact in this setting. The next theorem is the corresponding analogue for our definition of perfect equilibrium. It is a direct consequence of the previous results.

\begin{theorem} \label{theo: existence finadd}
Let $\Gamma$ be a strategic form game with finitely additive strategies. Then the set of perfect equilibria of $\Gamma$ is non-empty and compact.
\end{theorem}

\noindent
Proof. \quad
A strategic form game with finitely additive strategies satisfies the conditions of Theorem \ref{theorem-mainexistence}.
\hfill \qed
\vskip6pt

Our definition of perfect equilibrium is fully operational in the finitely additive setting. To illustrate this, we compute the set of perfect equilibria in the next example. The example also underscores the relevance of the finitely additive framework, as it provides an instance of a game with viable payoff functions $u_i$ for which the resulting game in countably additive strategies does not have a Nash equilibrium and, thus, no perfect equilibrium either.

\begin{example} \label{example: variant Wald game}
A variant of the Wald game. \rm
The game has two players, 1 and 2, and $A_1 = A_2 = \na$. The payoff functions $u_1$ and $u_2$ are symmetric and $u_1$ is given below. Player 1 is the row player and player 2 the column player.
$$
\begin{tabular}{|c|rrrc|}
\hline
$u_1$ & 1 & 2 & 3 & $\ldots$ \\ \hline
1 & 0 & 0 & 0 & $\ldots$ \\
2 & $\frac{1}{2}$ & 0 & 0 & $\ldots$ \\
3 & $\frac{1}{3}$ & $\frac{1}{3}$ & 0 & $\ldots$ \\

$\vdots$ & $\vdots$ & $\vdots$ & $\vdots$ & $\ddots$ \\
\hline
\end{tabular}
$$
Further, $\F$ is the collection of all subsets of $A = A_1 \times A_2 = \na \times \na$. The strategy sets are $\Xi_1 = \Xi_2 = \Xi(\na)$, and $\Xi = \Xi_1 \times \Xi_2$.
\vskip6pt

The game is a strategic form game in finitely additive strategies. The game played with countably additive strategies does not have Nash equilibria. In contrast, for the game played with finitely additive strategies, the set of Nash equilibria coincides with the set of pairs of diffuse charges, and all Nash equilibria are perfect. We have the following 4 Claims.
Detailed proofs are available in Appendix A of the Supplemental Appendix. 
\vskip6pt

{\bf Claim 1.} \quad
The payoff functions $u_1$ and $u_2$ are uniform limits of simple functions. Hence, the game is a strategic form game in finitely additive strategies.
\vskip6pt

{\bf Claim 2.} \quad
Let $(\kappa_1, \kappa_2) \in \Xi$ be a Nash equilibrium. Then there is a player $i$ with $\kappa_i(k) = 0$ for all $k \in \na$. In particular, the game does not have a Nash equilibrium in countably additive strategies.
\vskip6pt
 
{\bf Claim 3.} \quad
The set of Nash equilibria of this game is the collection of pairs of diffuse charges. Consequently, all Nash equilibria yield payoff zero for both players.
\vskip6pt

{\bf Claim 4.} \quad
The set of perfect equilibria of this game is the collection of pairs of diffuse charges.
\vskip6pt

{\bf Remark.}\quad
For a given game, the set of Nash equilibria and the set of perfect equilibria may well differ. See for instance Example 
\ref{example: PE is not NE}.
\hfill $\triangleleft$
\end{example}

\vspace{-.25cm}
\subsection{Games played with countably additive strategies}\label{subs_ca}

The second example of a classic class of games is the one where $A$ is metric and compact, $\Sigma = \Delta$ is endowed with the topology of weak convergence (and metrized by the Prohorov distance), and payoff functions are continuous. This class of games coincides with the one studied in \cite{Simon Stinchcombe:1995}.

\begin{definition}\label{def-strformgamecad} \rm
A strategic form game $\Gamma = (N, A, \F, \Delta, u)$ is called a \emph{strategic form game with countably additive strategies} if
\vspace{-.15cm}

\begin{itemize}
\item[{[1]}] Each $A_i$ is a compact metric space with some metric $d_i$.
\vspace{-.15cm}
\item[{[2]}] Each $u_i$ is continuous (and hence bounded).
\vspace{-.15cm}
\item[{[3]}] Each $\F_i$ is the associated Borel sigma-field.
\vspace{-.15cm}
\item[{[4]}] The topology on $\Delta$ is the topology of weak convergence, metrized by the Prohorov metric.
\end{itemize}
\end{definition}
\vspace{-.15cm}

In this case, since all other notions are well-known, we only specify the Prohorov metric in more detail.
We endow $A$ with the metric $d(a, b) = \max \{ d_i(a_i, b_i) \ \mid \ i \in N \}$.
Let $\F$ be the Borel sigma-field on $A$ induced by $d$.
For $\sigma, \tau \in \Delta$ the Prohorov metric $d_P(\sigma, \mu)$ is defined by
$$
d_P(\sigma, \tau) = \inf \left\{ \veps > 0 \ \mid \
\sigma(B) \le \tau(B_\veps) + \veps 
\hbox{ and }
\tau(B) \le \sigma(B_\veps) + \veps
\hbox{ for all } B \in \F
\right\},
$$
where, for every $B \in \F$ and $\veps > 0$,
$B_\veps = \left\{ a \in A \mid d(a, B) < \veps \right\}$.
\vskip6pt

We note that for a game with countably additive strategies, each $u_i$ is viable, and hence each expected payoff function $U_i$ exists. Moreover, each $U_i$ is multi-linear, and continuous in the Prohorov topology. These results are well-known, and follow for example from Theorem 2.8 in \cite{Billingsley: 1999}. A detailed proof is provided in Lemma \ref{theo: alpha is beta} in Appendix B of the Supplemental Appendix. We can now prove the following statement.

\begin{theorem} \label{theo: existence countably additive}
Let $\Gamma$ be a strategic form game with countably additive strategies. Then the set of perfect equilibria of $\Gamma$ is non-empty and compact.
\end{theorem}

\noindent
Proof. \quad
A strategic form game with countably additive strategies satisfies the conditions of Theorem \ref{theorem-mainexistence}.
\hfill \qed
\vskip6pt

The next example is due to \cite{Simon Stinchcombe:1995}. They show that the Nash equilibrium $(0, 0)$ is a weak perfect equilibrium, but not a strong perfect equilibrium. We show that $(0,0)$ is a perfect equilibrium as in Definition \ref{def-perf}.

\begin{example} 
A game with countably additive strategies. \quad \rm
Consider the following two-person strategic form game. $A_1 = \{ -1 \} \cup [0, 1]$ and $A_2 = [0, 1]$.
Further, $u_2(a_1, a_2) = - a_2$, and
$$
u_1(a_1, a_2) =
\begin{cases}
\hskip 8pt \frac{1}{8} \cdot a_2 & \hbox{if } a_1 = -1 \cr
\hskip22pt a_1 & \hbox{if } a_1 \in \left[ 0, \frac{1}{2} a_2 \right) \cr
a_2 - a_1 & \hbox{if } a_1 \in \left[ \frac{1}{2} a_2, 1 \right].
\end{cases}
$$
Simon and Stinchcombe show that the Nash equilibrium $(a_1, a_2) = (0, 0)$ is weak perfect according to their definition. We show that the equilibrium is perfect in our sense, which implies their result by Theorem \ref{theo: weak perfect}.
\vskip6pt

{\bf Claim.} \quad The strategy pair $(a_1, a_2) = (0, 0)$ is a perfect equilibrium.
\vskip6pt

Proof of claim. \quad
We prove item [2] of Theorem \ref{prop: characterization}. For each player $i$, take any carrier $\C_i$, and any open set $U_i \ni a_i$.
We argue that there are $\sigma_i, \kappa_i \in U_i$ such that $\C_i \subseteq \C(\sigma_i)$ and $\kappa_j$ is a best response to $\sigma_i$.
Take any strategy $\mu_i$ with carrier $\C_i$. Take any $\veps \in \left( 0, \frac{1}{8} \right]$. Take
$
\sigma_1 = (1 - \veps) \cdot \delta(0) + \veps \cdot \mu_1 \hbox{ and }
\sigma_2 = (1 - \veps) \cdot \delta(\veps) + \veps \cdot \mu_2.
$
Then $\sigma_i$ has at least carrier $\C_i$, and $\sigma_i \in U_i$ for sufficiently small $\veps > 0$.
Further, since player 2 has a dominant action $a_2 = 0$, we can take $\kappa_2 = \delta(0) \in U_2$.
\vskip6pt

Take any best response $\kappa_1$ for player 1 against $\sigma_2$. (Such a strategy exists due to compactness of $\Delta$ and continuity $U_i$.)  We argue that $\kappa_1$ puts weight one on the interval $[0, 2 \cdot \veps]$. If player 1 chooses $b_1 = -1$, we have that
$$
U_1(\delta(b_1), \sigma_2) = (1-\veps) \cdot U_1(\delta(b_1), \delta(\veps)) + \veps \cdot U_1(\delta(b_1), \mu_2) \le \frac{1 - \veps} {8} \cdot \veps + \frac{1}{8} \cdot \veps \le \frac{1}{4} \cdot \veps.
$$
If player 1 chooses $b_1 = \frac{1}{2} \cdot \veps$. Then, since $\veps \le \frac{1}{8}$, the payoff to player 1 is
\begin{eqnarray*}
U_1(\delta(b_1), \sigma_2)
& = &
(1-\veps) \cdot U_1(\delta(b_1), \delta(\veps)) + \veps \cdot U_1(\delta(b_1), \mu_2) \cr
& = &
(1 - \veps) \cdot \frac{1}{2} \cdot \veps + \veps \cdot
\left[ \int_0^\veps (b_2 - \frac{1}{2} \cdot \veps) d \mu_2 + \int_\veps^1 \frac{1}{2} \cdot \veps d \mu_2 \right] \cr
& \ge & 
(1 - \veps) \cdot \frac{1}{2} \cdot \veps +
\veps \cdot \left[ - \frac{1}{2} \cdot \veps + 0 \right] 
=\frac{1}{2} \cdot \veps \cdot (1 - 2 \cdot \veps) \ge \frac{3}{8} \cdot \veps.
\end{eqnarray*}
If $b_1 > 2 \cdot \veps$. Then $U_1(\delta(b_1), \delta(\veps)) = \veps - a_1$. Further, clearly $U_1(\delta(b_1), \mu_2) \le 1$. So,
$$
U_1(\delta(b_1), \sigma_2) = (1-\veps) \cdot U_1(\delta(b_1), \delta(\veps)) + \veps \cdot U_1(\delta(b_1), \mu_2) \le \veps - b_1 + \veps < 0.
$$
It follows that $0 \le b_1 \le 2 \cdot \veps$ for any pure best response $b_1$ against $\sigma_2$. So, $\kappa_1$ assigns probability 1 to the interval $[0, 2 \cdot \veps]$. Hence, $\kappa_1 \in U_1$ for sufficiently small $\veps > 0$.
\hfill $\triangleleft$
\end{example}



\section{Admissibility}\label{Section: Admissibility}

An admissible strategy is a strategy that satisfies some form of undominatedness.\footnote{For an elaborate motivation for admissibility
of Nash equilibrium refinements, we refer to \cite{Kohlberg Mertens: 1986}. In that paper, a Nash equilibrium is admissible if no player uses a weakly dominated strategy.}
In this section, we analyze two versions of admissibility, namely dominance (Thm \ref{theo: dominant is perfect}) and limit undominatedness (Thm \ref{theo: limit undominated}).
\vskip6pt

Let $\Gamma = (N, A, \F, \Sigma, U)$ be a strategic form game. A strategy is \emph{(weakly) dominant} if it is a best response against any strategy profile.\footnote{In this paper we only discuss weak dominance. The adjective "weak" is therefore omitted.}
A strategy $\sigma_i \in \Sigma_i$ is (weakly) dominated by a strategy $\tau_i \in \Sigma_i$ if
for all $\rho \in \Sigma$ it holds that
$
U_i(\sigma_i \mid \rho) \le U_i(\tau_i \mid \rho)
$
where the inequality is strict for at least one strategy profile $\rho \in \Sigma$.
A strategy $\sigma_i \in \Sigma_i$ is \emph{undominated} if it is not dominated by any strategy $\tau_i \in \Sigma_i$.
A strategy $\sigma_i \in \Sigma_i$ is \emph{limit undominated} if there is a net $(\sigma_{\alpha, i})_{\alpha \in I}$ of undominated strategies $\sigma_{\alpha, i} \in \Sigma_i$ that converges to $\sigma_i$. A strategy profile $\sigma = (\sigma_i)_{i \in N}$ is (limit) undominated if each $\sigma_i$ is (limit) undominated.

\vspace{-.25cm}
\subsection{Discussion of previous notions of admissibility}

There are games with infinitely many actions featuring a unique Nash equilibrium, in which all players use a weakly dominated strategy (see for example \cite{Simon Stinchcombe:1995}). The next example illustrates this. It motivates the use of limit undominatedness over that of undominatedness.
\vskip6pt

\begin{example} \quad
\label{example: another Wald game}
Game with unique Nash equilibrium in weakly dominated strategies. \rm

The game has two players, 1 and 2, with action sets $A_1 = A_2 = \na \cup \{ \infty \}$. The payoff functions $u_1$ and $u_2$ are symmetric, and $u_1$ is given by
$$
\begin{tabular}{|c|rrrcl|}
\hline
$u_1$ & 1 & 2 & 3 & $\ldots$ & $\infty$ \\ \hline
1 & 0 & 0 & 0 & $\ldots$ & 0 \\
2 & $\frac{1}{2}$ & 0 & 0 & $\ldots$ & 0 \\
3 & 0 & $\frac{1}{3}$ & 0 & $\ldots$ & 0 \\
$\vdots$ & $\vdots$ & $\vdots$ & $\vdots$ & $\ddots$ & $\vdots$ \\
$\infty$ & 0 & 0 & 0 & $\ldots$ & 0 \\
\hline
\end{tabular}
$$
Player 1 is the row player, and player 2 is the column player. The strategy spaces are $\Delta(A_1)$ and $\Delta(A_2)$. So, this is a game played with countably additive strategies. The only Nash equilibrium is $(\infty, \infty)$.\footnote{
It is clear that $(\infty, \infty)$ is a Nash equilibrium. To prove uniqueness, let $(\sigma_1, \sigma_2)$ be any strategy pair for which there are a player $i$ and an action $k_i$ with $\sigma_i(k_i) > 0$. Let $\ell_i$ be the smallest natural number on which player $i$ puts strictly positive weight. Then $\sigma_j(k_j) > 0$ for some $k_j > \ell_i$. Therefore player 1 can strictly improve his payoff by moving the probability $\sigma_1(\ell_1)$ from $\ell_1$ to $k_j + 1$. Hence, $(\sigma_1, \sigma_2)$ is not a Nash equilibrium.}
Then by Theorem \ref{theo: existence countably additive} it is the only perfect equilibrium. But, action $\infty$ is weakly dominated by any other action $a \ge 2$.
\hfill $\triangleleft$
\end{example}

To address this issue \cite{Simon Stinchcombe:1995} required a Nash equilibrium only to be limit admissible.\footnote{In \cite{Simon Stinchcombe:1995}, a Nash equilibrium is limit admissible if each player uses a strategy that is a limit of strategies that are admissible, where a strategy is called admissible if the strategy puts zero probability on weakly dominated actions.}
However, admissibility and limit admissibility as defined in that paper are fairly permissive. The next example demonstrates this. 

\begin{example} \label{example: dominated NE} \quad
A game with an admissible Nash equilibrium in which both strategies are weakly dominated. \rm
\quad
The game is a $3 \times 3$ bimatrix game. The payoff matrices for player 1 and 2 are as follows.
$$
u_1 =
\left[
\begin{matrix}
4 & 0 & 0 \cr
0 & 4 & 0 \cr
2 & 2 & 1
\end{matrix}
\right]
\quad \hbox{and} \quad
u_2 =
\left[
\begin{matrix}
4 & 0 & 2 \cr
0 & 4 & 2 \cr
0 & 0 & 1
\end{matrix}
\right].
$$
The strategy pair $\bigl((\frac{1}{2}, \frac{1}{2}, 0), (\frac{1}{2}, \frac{1}{2}, 0)\bigr)$, putting weight $\frac{1}{2}$ on the first two rows, resp.\ columns, is a Nash equilibrium. It is admissible in the sense of Simon and Stinchcombe, since both strategies put strictly positive weight only on actions that are not dominated by any strategy. However, for both players, the strategy $(\frac{1}{2}, \frac{1}{2}, 0)$ is weakly dominated by $(0, 0, 1)$. In particular, the equilibrium is not perfect.
\hfill $\triangleleft$
\end{example}

Thus, admissibility and limit admissibility are fairly permissive concepts. This paper proposes to use the more restrictive notion of limit undominatedness.

\vspace{-.25cm}
\subsection{Dominant strategy Nash equilibria are perfect}

This section shows that, for every strategic form game, each dominant strategy Nash equilibrium is a perfect equilibrium.

\begin{theorem} \label{theo: dominant is perfect}
Let $\Gamma = (N, A, \F, \Sigma, U)$ be a strategic form game as in Definition \ref{def-strformgame}. Then every profile of (weakly) dominant strategies in $\Gamma$ is a perfect equilibrium of $\Gamma$.
\end{theorem}

\noindent
Proof. \quad
Let $\kappa = (\kappa_i)_{i \in N}$ be a profile of dominant strategies. Then $\kappa$ is a Nash equilibrium. We show that $\kappa$ is perfect. Take any carrier $\C$, and any $\rho \in \Sigma$ with $\C(\rho) = \C$. Let, for any $k \in \na$,
$
\sigma^k = \frac{k-1}{k} \cdot \kappa + \frac{1}{k} \cdot \rho,
$
and $\kappa^k = \kappa$. Then, since addition and scalar multiplication are continuous in a topological vector space, $\sigma^k \rightarrow \kappa$ as $k \rightarrow \infty$. Further, $\kappa^k \rightarrow \kappa$ as $k \rightarrow \infty$, and for each $k$, $\kappa^k$ is a best response to $\sigma^k$. Hence, $\kappa$ is perfect by Theorem \ref{prop: characterization}.
\hfill \qed
\vskip6pt

The above result thus even applies to games that are outside the context of the classes of games discussed in Section \ref{Section: Standard frameworks}, as the next Example illustrates.
\vskip6pt

\begin{example} The Wald game. \rm
The game is played by two players with action spaces $A_1 = A_2 = \na$. When player 1 chooses $m \in \na$ and player 2 chooses $n \in \na$, the player who chose the higher number receives payoff 1, while the other player receives payoff 0. In case of a draw, both players receive zero. Note that the payoff functions are not limits of simple functions, so that the game does not fall into the classic setting.
\vskip6pt

We consider the case where $\Sigma_1 = \Sigma_2 = \Xi(\na)$. A diffuse charge wins from any action. If both players choose a diffuse charge, the payoffs are defined as follows: player 1 gets an amount $\alpha \in [0, 1]$, and player 2 gets $1 - \alpha$.\footnote{A version of this Example is also discussed in Example 2.1 of \cite{Stinchcombe: 2005} and in Examples 3.1, 6.1, and 6.2 of \cite{Zseleva: 2017}.}
We prove the following claim.
\vskip6pt

{\bf Claim.} \quad
In this game, [1] a strategy is dominant precisely when it is a diffuse charge, and [2] a strategy pair $\kappa = (\kappa_1, \kappa_2)$ is a Nash equilibrium precisely when $\kappa_1$ and $\kappa_2$ are diffuse charges. Consequently, any Nash equilibrium is perfect.
\vskip6pt

Proof: Claim [1] is straightforward to check. In order to prove Claim [2] note that, by Theorem \ref{theo: dominant is perfect}, any pair of diffuse charges is a perfect equilibrium. Thus,  we only need to verify that $\kappa$ is a Nash equilibrium only when $\kappa_1$ and $\kappa_2$ are diffuse charges.
\vskip6pt

Suppose $\kappa = (\kappa_1, \kappa_2)$ is a Nash equilibrium and that $\kappa_1(k) > 0$ for some $k \in \na$. Then $\kappa_2$ puts zero weight on the set $\{1, 2, \ldots, k \}$. Since $\kappa_1(k) > 0$, $\kappa_1$ is not a best response to $\kappa_2$, contradicting the assumption that $\kappa$ is a Nash equilibrium.
It follows that $\kappa_1(k) = \kappa_2(k) = 0$ for all $k \in \na$. Hence both $\kappa_1$ and $\kappa_2$ are diffuse charges.
\hfill $\triangleleft$
\end{example}

\vspace{-.25cm}
\subsection{Results on admissibility for friendly games} \label{subsec: limit undominated}

In this section we identify a class of games in which each perfect equilibrium is limit undominated. We coin the elements of this class friendly games. Within the class of friendly games we also show that, if a player has a dominant strategy, that player uses a dominant strategy in every perfect equilibrium.

\vspace{-.15cm}
\subsubsection*{Friendly games, definition and examples}

The strategic form game $\Gamma = (N, A, \F, \Sigma, u)$ is a \emph{friendly game} when
\begin{itemize}
\vspace{-.15cm}
\item each $A_i$ is endowed with a topology $T_i$ such that $(A_i, T_i)$ is a Hausdorff space,
\vspace{-.15cm}
\item each $\F_i$ is the Borel field associated with $T_i$,
\vspace{-.15cm}
\item each $u_i$ is continuous, and the uniform limit of simple functions, and
\vspace{-.15cm}
\item each $\Sigma_i$ contains a strictly positive strategy, where a strategy $\rho_i \in \Sigma_i$ is called \emph{strictly positive} if $T_i \setminus \{ \emptyset \} \subseteq \C(\rho_i) $.\footnote{In the literature on measure theory strictly positive strategies are often called strictly positive measures. See for example \cite{Argyros: 1983}, and \cite{Fremlin 2001} in remark 411N-(f).}
 In other words, strategy $\rho_i$ assigns strictly positive probability to each non-empty open set.
\end{itemize}

\begin{remark} \rm
Since each $u_i$ is the uniform limit of simple functions, each $u_i$ is viable, so that, by Lemma \ref{viable is multilinear} in Appendix A of the Supplemental Appendix, $U_i$ exists and $U_i$ is multilinear.
\end{remark}

\begin{example} Friendly games. \rm \label{example: friendly spaces}
A strategic form game $\Gamma$ with countably additive strategies as in Definition \ref{def-strformgamecad} is friendly. A strategic form game $\Gamma = (N, A, \F, \Sigma, u)$ with finitely additive strategies as in Definition \ref{def-strformgamefinad}, where each $(A_i, T_i)$ is second countable, is friendly. For example, when each player has action space $\na$ with the discrete topology. Thus, the viable versions of the Wald game in Examples \ref{example: variant Wald game} and \ref{example: another Wald game} are friendly games.
\hfill $\triangleleft$
\end{example}

\begin{example} \label{example: friendly spaces 2}
A friendly game that is not metrizable. \rm
Let $\Gamma = (N, A, \F, \Xi, u)$ be a strategic form game with $A_i = [0, 1]$ for each $i \in N$. Then the uniform probability measure $\lambda \in \Delta(A_i)$ is strictly positive. So, $\Xi_i = \Xi(A_i)$ admits a strictly positive strategy. But, $\Xi_i$ is not second countable, hence not metrizable. To see this, for each $x \in A_i$, take
$
V_x =
\left\{ \kappa_i \in \Xi_i \mid \kappa_i(\{ x \}) > \frac{1}{2} \right\}.
$
There are uncountably many sets $V_x$, and each $V_x$ is open in Tychonov topology. It suffices to argue that the sets $V_x$ are mutually disjoint.
Take any $x, y \in A_i$ with $y \not= x$, and any $\kappa_i \in V_x$. Then $\kappa_i(x) > \frac{1}{2}$. By additivity it follows that $\kappa_i(y) < \frac{1}{2}$ for any $y \not= x$, so that $\kappa_i \notin V_y$.
\hfill $\triangleleft$
\end{example}

\begin{example} An unfriendly game. \rm
\label{example: not friendly}
Absence of strictly positive strategies may occur even when the action space is compact. To see this, reconsider Example \ref{example: friendly spaces 2}. We already observed that $\Xi_i$ is compact, but admits an uncountable collection of mutually disjoint open sets $V_x$. So, for a game $\Gamma = (N, B, \F, \Xi, u)$ with $B_i = \Xi_i(A_i)$ for some $i \in N$, the strategy set $\Xi(B_i)$ does not contain a strictly positive strategy.
\hfill $\triangleleft$
\end{example}

\subsubsection*{Dominance and limit undominatedness for friendly games}

This subsection generalizes Theorem 5.1 in \cite{Elnaz: 2013} on dominance and limit undominatedness. For a friendly game, each perfect equilibrium is limit undominated. Moreover, if a player has a dominant strategy, in every perfect equilibrium that player uses such a strategy.

\begin{theorem} \label{theo: limit undominated}
Let $\Gamma = (N, A, \F, \Sigma, u)$ be a friendly game. Then every perfect equilibrium is limit undominated.
Moreover, if a player $i$ has a dominant strategy and $U_i$ is continuous, then in every perfect equilibrium player $i$ uses a dominant strategy.
\end{theorem}

\noindent
Proof.\quad
{\bf A.} \quad
Let $\kappa$ be a perfect equilibrium. We show that $\kappa$ is limit undominated.
\vskip6pt

\noindent
Since the game is friendly, for each $i$ we can take a strictly positive strategy $\gamma_i \in \Sigma_i$. Write $\C = (\C_i(\gamma_i))_{i \in N}$. Take nets $(\sigma_\alpha)_{\alpha \in I}$ and $(\kappa_\alpha)_{\alpha \in I}$ as in Definition \ref{def-perfstr} of a perfect strategy profile. Consider the nets $(\sigma_\alpha)_{\alpha \in I}$ and $(\kappa_\alpha)_{\alpha \in I}$, restricted to those indices $\alpha$ with $\C(\sigma_\alpha) \supseteq \C$. For each $\alpha$, since $\kappa_{\alpha,i}$ is a best response to $\sigma_\alpha$ it follows from Lemma \ref{theo: undominated BRs} that $\kappa_{\alpha,i}$ is undominated. Then $\kappa_{\alpha,i}$ is a net of undominated strategies that converges to $\kappa_i$. Hence, $\kappa_i$ is limit undominated.
\vskip6pt

{\bf B.} \quad
Suppose that $U_i$ is continuous and player $i$ has a dominant strategy. Let $\kappa$ be a perfect equilibrium. We show that $\kappa_i$ is dominant.
\vskip6pt

Since $\kappa$ is a perfect equilibrium, part A implies that $\kappa_i$ is the limit of undominated strategies of player $i$.
Further, since player $i$ has a dominant strategy, any undominated strategy of player $i$ is dominant. It follows that $\kappa_i$ is the limit of dominant strategies of player $i$.
The limit of dominant strategies is dominant because $U_i$ is continuous.
\hfill \qed
\vskip6pt

{\bf Remark.} \quad
We do not know whether Theorem \ref{theo: limit undominated} is also valid for games where at least one player does not have a strictly positive strategy, as in Example \ref{example: not friendly}. 

\section{Invariance} \label{Section: Invariance}

The notion of invariance for Nash equilibrium dates back at least to the papers by \cite{Thompson: 1952} and \cite{Dalkey: 1953}. They show that for two extensive form games, of which one can be transformed into the other by using certain basic inessential transformations, the sets of Nash equilibria of those two games coincide (up to the transformation of one game into the other).  \cite{Kohlberg Mertens: 1986} argue that any reasonable refinement of Nash equilibrium should also satisfy invariance under such transformations. \cite{Mertens: 1987} even argues that any reasonable solution concept should satisfy the more demanding notion of ordinality, which implies both invariance and admissible best reply invariance.\footnote{This requires two games with the same admissible best reply correspondence to have identical solutions.}
\vskip6pt

Theorem \ref{perfect equilibrium is invariant} shows that, for strategic form games with finitely additive strategies, perfect equilibrium satisfies a weak form of the invariance requirement. Nevertheless, we also show that payoff equivalent strategies need not be interchangeable (cf. Example \ref{example: hazy filters}). It remains an open question under what conditions stronger versions of invariance are met by our notion of perfect equilibrium.
\vskip6pt

Example \ref{example: PE not invariant} (due to \cite{Elnaz: 2013}) shows that, for strategic form games with countably additive strategies, perfect equilibrium violates the weak form of invariance. Example \ref{example: hazy filters} analyses the same game, but now allows players to use finitely additive strategies. Computing the set of perfect equilibria shows that this indeed restores invariance. The example also illustrates that stronger forms of invariance are violated for games with finitely additive strategies.

\vspace{-.25cm}
\subsection{The definition of invariance}

Invariance of solution concepts is closely linked to the identification of payoff equivalent strategy profiles. We first discuss these notions.
\vskip6pt

Let $\Gamma = (N, A, \F, \Xi(A), u)$ and $\Gamma^\prime =(N, B, \G, \Xi(B), v)$ be two games with finitely additive strategies as in Definition \ref{def-strformgamefinad}.
Let $\varphi_i \colon A_i \rightarrow B_i$ be a surjective map with
$
\phi_i^{-1}(G_i) \in \F_i \quad \hbox{for every } G_i \in \G_i$, i.e., $\varphi_i$ is measurable with respect to $\F_i$ and $\G_i$. 
A measurable surjective map $\varphi_i$ induces a map $\Phi_i \colon \Xi(A_i) \rightarrow \Xi(B_i)$ by
$
\Phi_i(\kappa_i)(G_i) = \kappa_i\left( \varphi_i^{-1}(G_i) \right)
$
for all $\kappa_i \in \Xi(A_i)$ and $G_i \in \G_i$. Define $\varphi \colon A \rightarrow B$ by
$
\varphi(a) = (\varphi_i(a_i))_{i \in N}
$
for all $a = (a_i)_{i \in N} \in A$. We say that $\varphi \colon A \rightarrow B$ \emph{respects payoffs} if $u_i(a) = v_i(\varphi(a))$ for all $a \in A$ and all $i \in N$.\footnote{Two action profiles $a \in A$ and $b \in A$ are payoff equivalent if $u_i(a) = u_i(b)$ for every player $i \in N$. Note that, if $\phi$ respects payoffs, then $\phi(a) = \phi(b)$ implies that $a$ and $b$ are payoff equivalent.}
We define $\Phi \colon \Xi(A) \rightarrow \Xi(B)$ by
$
\Phi(\kappa) =  (\Phi_i(\kappa_i))_{i \in N}.
$
A \emph{solution concept} is a correspondence $S$ that assigns to each game $\Gamma = (N, A, \F, \Xi(A), u)$ with finitely additive strategies a set $S(\Gamma) \subseteq \Xi(A)$ of strategy profiles.\footnote{Formally speaking, invariance requires specification of the universe of games on which invariance comparisons between games are being considered. However, since we only consider invariance on the class of strategic form games with finitely additive strategies, this is the context in which the invariance requirement is formulated in this paper.}

\begin{definition}\label{def-invariant} \rm
A solution concept $S$ is called \emph{invariant} if for every two games $\Gamma = (N, A, \F, \Xi(A), u)$ and $\Gamma^\prime =(N, B, \G, \Xi(B), v)$ and every $\varphi \colon A \rightarrow B$ that respects payoffs it holds that
$
\Phi(S(\Gamma)) \subseteq S(\Gamma^\prime).
$
\end{definition}

\vspace{-.25cm}
\subsection{Invariance of perfect equilibrium}

We show that for strategic form games with finitely additive strategies, the perfect equilibrium solution concept satisfies invariance.

\begin{theorem} \label{perfect equilibrium is invariant}
Let $\Gamma = (N, A, \F, \Xi(A), u)$ and $\Gamma^\prime =(N, B, \G, \Xi(B), v)$ be two games with finitely additive strategies as in Definition \ref{def-strformgamefinad}. Let $\varphi \colon A \rightarrow B$ be a map that respects payoffs. If $\kappa \in \Xi(A)$ is a perfect equilibrium of $\Gamma$, then $\Phi(\kappa)$ is a perfect equilibrium of $\Gamma^\prime$.
\end{theorem}

Proof. \quad
Take a perfect equilibrium $\kappa \in \Xi(A)$ in $\Gamma$. We show that $\eta = \Phi(\kappa)$ is a perfect equilibrium in $\Gamma^\prime$. It suffices to prove [2] of Theorem \ref{prop: characterization}.
Take any carrier $\D \subseteq \G$ and any subbasis element
$
V = \Bigl\{ \zeta \in \Xi(B) \mid
\left| \eta(G) - \zeta(G) \right| < \veps \Bigr\},
$
where $G \in \G$ is fixed. Take $\mu \in \Xi(B)$ with $\C(\mu) = \D$.
By Lemma \ref{lemm: Phi surjectie} we can take $\zeta \in \Xi(A)$ with $\Phi(\zeta) = \mu$. Write $\C = \C(\zeta)$ and $U = \Phi^{-1}(V)$. Then, since $\Phi(\kappa) = \eta \in V$, we know that $\kappa \in U$. So, since $\kappa$ is perfect, by [2] of Theorem \ref{prop: characterization} we can take $\nu \in U$ and $\xi \in U$ with $\C(\nu) \supseteq \C$ and $\xi$ is a best response to $\nu$. It follows that $\Phi(\xi) \in V$, $\Phi(\nu) \in V$, and, since $U_i(\kappa) = V_i(\Phi(\kappa))$
for all $\kappa$ by the definition of expected payoff, $\Phi(\xi)$ is a best response to $\Phi(\nu)$.
\vskip6pt

Moreover, take any $D \in \D = \C(\mu)$. Then
$
\zeta\bigl(\phi^{-1}(D)\bigr) = \Phi(\zeta)(D) = \mu(D) > 0.
$
So, $\phi^{-1}(D) \in \C(\zeta) = \C$. So, since $\C(\nu) \supseteq \C$,
$
\Phi(\nu)(D) = \nu\bigl(\Phi^{-1}(D)\bigr) > 0.
$
It follows that $\C(\Phi(\nu)) \supseteq \D$.
\hfill \qed

\vspace{-.25cm}
\subsection{Illustrative examples}

We illustrate the above result with several examples. The next example demonstrates how the result can facilitate identification of perfect equilibrium.

\begin{example} 
Perfect equilibrium and invariance. \rm
Let $\Gamma = (N, A, \F, \Xi, u)$ be the strategic form game with finitely additive strategies, where $N$, $A_1$, $A_2$, $\F_1$, $\F_2$, $u_1$, and $u_2$ are as in Example A of Section \ref{Section: Examples}. The reduced form\footnote{If a map $\varphi$ that respects payoffs exists from $\Gamma$ to $\Gamma^\prime$, and moreover for any two payoff equivalent strategy profiles $a$ and $c$ it holds that $\varphi(a) = \varphi(c)$, we say that $\Gamma^\prime$ is the \emph{reduced form} of $\Gamma$.}
$\Gamma^\prime$ of the game $\Gamma$ is the $2 \times 2$ bimatrix game
$$
\Gamma^\prime =
\left[ \ 
\begin{matrix}
0, 0 & 0, 0 \cr
0, 0 & 1, 1
\end{matrix} \ 
\right].
$$
The action spaces are $A_1 = \{ U, D \}$ and $A_2 = \{ L, R \}$. In the finitely additive case where $\Sigma = \Xi$, it follows from Theorem \ref{perfect equilibrium is invariant} that $\Phi(PE(\Gamma))$ is a subset of $\{ (D, R) \}$. So, $PE(\Gamma)$ is a subset of
$
\Phi^{-1}\bigl((D, R)\bigr) = \bigl\{(\delta(1), \delta(1)) \bigr\}.
$
Now note that the payoff functions of $\Gamma$ are simple functions. So, $\Gamma$ has at least one perfect equilibrium by Theorem \ref{theo: existence finadd}. So, we can conclude that $PE(\Gamma) = \{(\delta(1), \delta(1)) \}$.
\vskip6pt

For $\Sigma = \Delta$, the map $\Phi$ is not continuous. Nevertheless, as argued already in Example A of Section \ref{Section: Examples}, also in this case $PE(\Gamma) = \{(\delta(1), \delta(1)) \}$.
\hfill $\triangleleft$
\end{example}

The set of perfect equilibria for a game played with countably additive strategies may differ from the set of perfect equilibria for the same game played with finitely additive strategies. In general this is clear, since for example a perfect equilibrium in diffuse strategies does not exist in the game played with countably additive strategies. The next Example illustrates that this is true even for the set of pure Nash equilibria. It is identical to Example 6 in \cite{Elnaz: 2013}.
 
\begin{example} \label{example: PE not invariant}
Non-invariant Perfect equilibrium. \rm
The game is $\Gamma = (N, A, \F, \Delta, u)$, where $N = \{ 1, 2\}$, and the action spaces are $A_1 = \{ U, D \}$ and $A_2 = \{ 1, 2, 3, 4, \ldots, \infty \}$, where all natural numbers are isolated points, whereas $\infty$ is the limit point of the sequence $1, 2, 3, \ldots$. Further, $\F_1 = \F_2 = 2^{\na}$, and the payoffs are given by
$$
u_1 =
\left[
\begin{matrix}
\hfill 0 & \hfill 0 & \hfill 0 & \hfill 0 & \ldots\ldots & 0 \ \ \ \cr \cr
1 & 1 & 1 & 1 & \ldots\ldots & 1\ \ \ 
\end{matrix}
\right]
\quad \hbox{and} \quad
u_2 =
\left[
\begin{matrix}
\hfill 0 & \hfill \frac{1}{2} & \hfill \frac{1}{3} & \hfill \frac{1}{4} & \ldots\ldots & 0 \ \ \ \cr \cr
0 & - \frac{1}{4} & - \frac{1}{9} & - \frac{1}{16} & \ldots\ldots & 0\ \ \ 
\end{matrix}
\right].
$$
As \cite{Elnaz: 2013} observe, $(D, 1)$ is not a weak perfect equilibrium. Hence, by Theorem \ref{theo: weak perfect} $(D, 1)$ is not a perfect equilibrium either. But, as we show below, $(D, \infty)$ is perfect, even though the actions $1$ and $\infty$ are payoff equivalent.

\begin{claim}
The strategy pair $(D, \infty)$ is the unique perfect equilibrium of the game $\Gamma$.
\end{claim}

Proof of claim. \quad
In the countably additive case, let $(\sigma_1, \sigma_2)$ be a perfect equilibrium. Then, since $D$ is the dominant strategy, $\sigma_1 = \delta(D)$. Against this strategy, in any Nash equilibrium player 2 will only put strictly positive weight on $1$ or $\infty$. However, in Example 6 of \cite{Elnaz: 2013} it is shown that $\delta(1)$ is not a best response against any full support strategy of player 1. The result now follows from Theorem \ref{theo: existence countably additive} on existence of perfect equilibrium. End proof of claim.
\hfill $\triangleleft$
\end{example}

\begin{example} \label{example: hazy filters}
Perfect equilibrium in diffuse charges. \rm
We reconsider the game in the previous Example \ref{example: PE not invariant}, and analyse it when we replace $\Delta$ by $\Xi$. We denote this game by $\Gamma^\ast$. As it turns out, and as is to be expected in the light of the previous discussion, invariance is restored. However, the way in which invariance is restored is worth emphasizing. In the game $\Gamma^\ast$, neither $(B, 1)$ nor $(B, \infty)$ are perfect, and in any perfect equilibrium in finitely additive strategies, player 2 only uses diffuse charges. Thus, in this example, the set of perfect equilibria of $\Gamma$ is disjoint from the set of perfect equilibria of $\Gamma^\ast$. 
Moreover, although all pairs of diffuse charges are payoff equivalent, not every pair of diffuse charges is perfect. This is a remarkable difference with the situation for games with finite action sets.
\vskip6pt

We fully characterize the set of perfect equilibria. The perfect equilibrium strategies for player 2 are a specific, non-trivial, class of diffuse charges. We call such strategies twinned hazy filters. We define the notion of twinned hazy filter below. 
\vskip6pt

A diffuse charge $\kappa$ is called an \emph{ultrafilter}\footnote{Slight abuse of terminology. In fact $\kappa$ is the indicator function associated with a free ultrafilter.}
if $\kappa(E) \in \{0, 1\}$ for all $E \subseteq \na$.
A diffuse charge $\kappa$ is called a \emph{hazy filter} if there are ultrafilters $\eta$ and $\xi$ with
$
\kappa = c \cdot \eta + (1 - c) \cdot \xi
$
for some $c \in (0, 1)$. The following characterization provides a useful interpretation for the notion of a hazy filter. A detailed proof is available in Appendix E of the Supplemental Appendix.
\vskip6pt

{\bf Claim A.} \quad
Let $\kappa$ be a diffuse charge. Then $\kappa$ is a hazy filter precisely when for any partition $E_1, E_2, E_3$ of $\na$ it holds that $\kappa(E_i) = 0$ for at least one $i = 1, 2, 3$.
\vskip6pt

Two ultrafilters $\eta$ and $\xi$ are called \emph{twins} if
for every two subsets $X$ and $Y$ of $\na$ with $\eta(X) = \xi(Y) = 1$ there is a subset $B_k = \{ k, k+1 \}$ of $\na$ such that both $X \cap B_k$ and $Y \cap B_k$ are not empty.
A hazy filter $\kappa = c \cdot \eta + (1 - c) \cdot \xi$ is called \emph{twinned} if $\eta$ and $\xi$ are twins. Note that $\eta = \xi$ is allowed. Thus, any ultrafilter is a twinned hazy filter. 
\vskip6pt

{\bf Claim B.} \quad
Let $\kappa = (\kappa_1, \kappa_2)$ be a strategy pair. Then $\kappa$ is a perfect equilibrium of $\Gamma$ precisely when $\kappa_1 = \delta(D)$ and $\kappa_2$ is a twinned hazy filter.
\vskip6pt

A detailed proof is available in Appendix E of the Supplemental Appendix.
\hfill $\triangleleft$
\end{example}

{\bf Remark.} \quad
Note that all diffuse charges are payoff equivalent. Yet, not all diffuse charges are perfect equilibrium strategies.
For example, let $E_1$, $E_2$, and $E_3$ be a partition of $\na$ into three infinite sets, and let $\kappa_i$ be an ultrafilter with $\kappa_i(E_i) = 1$. Define
$$
\kappa = \frac{1}{3} \cdot (\kappa_1 + \kappa_2 + \kappa_3).
$$
Then $\kappa$ is a diffuse charge. However, since $\kappa(E_j) > 0$ for $j=1, 2, 3$, we know by Claim A that $\kappa$ is not a hazy filter.

\section{Relations to earlier notions of perfection} \label{Section: Earlier Notions}

This section relates our notion of perfection to earlier such notions in the literature.
In the case where each player has finitely many actions, we show that a strategy profile is a perfect equilibrium according to Definition \ref{def-perf} precisely when it is a perfect equilibrium according to the definition of \cite{Selten: 1975}.
\vskip6pt

For strategic form games with countably additive strategies, we show that any equilibrium that is perfect in our sense is also a weak perfect equilibrium according to \cite{Simon Stinchcombe:1995}. For the case where the game has two players, and one of the two players has two pure actions, we show that the two notions do indeed coincide. 
\vskip6pt

For strategic form games with finitely additive strategies we show that an equilibrium that is perfect according to the definition in \cite{Marinacci: 1997} (M-perfect) is also perfect according to our definition.\footnote{The set of M-perfect equilibria is empty for games in which at least one player has an uncountable action set. Hence, the result is trivial in those cases.}
For strategic form games with finitely additive strategies in which each player has at most countably many actions, we show that the two notions coincide for Nash equilibria in countably additive strategies.

\vspace{-.25cm}
\subsection{Games with finite action spaces}\label{sec-fin}

The analysis in this section focuses on games with finite action spaces. Consider a strategic form game $\Gamma = (N, A, \F, \Sigma, u)$ as defined in Definition \ref{def-strformgame} and restrict the set $A_i$ of actions to be finite for each player $i \in N$.
Then, for each player $i\in N$, the set $\Delta_i = \Xi_i$ of strategies with its usual topology is compact and metrizable. Further, $\mathcal F_i = 2^{A_i}$, and there is a maximal carrier $\mathcal C_i=2^{A_i}\setminus\{\emptyset\}$.
\vskip6pt

\cite{Selten: 1975} defined perfect equilibrium in this context of games with finite action sets. For games with finite action sets, the literature knows several alternative definitions (cf. van Damme [1991]). We use the following formulation.
\vskip6pt

A strategy profile $\sigma$ is a perfect equilibrium if there exists a sequence of strategy profiles $(\sigma^k)_{k=1}^\infty$ such that {[D1]} $\sigma^k$ has maximal carrier for each $k\in \N$, and {[D2]} $\sigma^k \rightarrow \sigma$ as $k \rightarrow \infty$, and {[D3]} $\sigma \in \text{BR}(\sigma^k)$ for each $k\in \N$.

\begin{theorem} \label{theo: finite games}
Suppose that the game has finite action spaces. Then,  a strategy profile $\sigma$ is a perfect equilibrium as in Definition \ref{def-perf} precisely when $\sigma$ is perfect according to the above mentioned definition.
\end{theorem}

Proof. \quad
First, assume that $\sigma$ is a perfect equilibrium in the above sense. Then, there exists a sequence of strategy profiles $(\sigma^k)_{k=1}^\infty$ satisfying the conditions [D1], [D2] and [D3]. Let $\kappa^k=\sigma$ for each $k\in\N$. Then, by [D1] the sequence $(\sigma^k)_{k=1}^\infty$ is a completely mixed net, and the conditions of Definition \ref{def-perfstr} translate into [1'] $\sigma^k\to\sigma$ as $k\to\infty$, [2'] $\kappa^k \to \sigma$ as $k\to\infty$, and [3'] $\kappa^k=\sigma\in \BR(\sigma^k)$ for each $k\in\N$. By [D2] and [D3], these conditions are satisfied.
\vskip6pt

Now assume that $\sigma$ is a perfect equilibrium as in Definition \ref{def-perf}. Take $\C_i = 2^{A_i}\setminus\{\emptyset\}$. Since the game has finite action spaces, we can take an open set $U_i \ni \sigma_i$ such that $\C(\tau_i) \supseteq \C(\sigma_i)$ for any $\tau_i \in U_i$. For $k \in \na$, write
$
V_{i,k} = \left\{ \tau_i \in U_i \mid \| \tau_i - \sigma_i \| < \frac{1}{k} \right\}.
$
Write $\C = \prod_i \C_i$ and $V_k = \prod_i V_{i,k}$.
Then according to Theorem \ref{prop: characterization} there are $\tau^k, \kappa^k \in V_k$ with $\C(\tau^k) = \C$ and $\kappa^k$ is a best response to $\tau^k$. Then, $\tau^k \to \sigma$ as $k\to\infty$ ([D2]). Also, since $\C(\tau) = \C$ we know that $\tau^k$ has maximal carrier for each $k\in \N$ ([D1]). Third, $\kappa^k$ is a best response to $\tau^k$. Then however, since $\C(\kappa^k) \supseteq \C(\sigma)$, also $\sigma$ is a best response to $\tau^k$ ([D3]).
Thus, conditions [D1], [D2], and [D3] are satisfied for the sequence $(\tau^k)_{k = 1}^\infty$. Hence, $\sigma$ is a perfect equilibrium in the above sense.
\hfill \qed

 \vspace{-.25cm}
\subsection{Games with countably additive strategies}

This section focuses on the classic setting of games with countably additive strategies. Theorem \ref{theo: weak perfect} shows that every equilibrium that is perfect in our sense is weakly perfect in the sense of \cite{Simon Stinchcombe:1995}. Theorem  \ref{theo: two players} shows that the two notions coincide when there are only two players, and at least one player only has two actions.  
\vskip6pt

{\eightrm A GENERAL RESULT} \quad
Following \cite{Simon Stinchcombe:1995}, we define weak perfect equilibrium as follows.\footnote{The original definition differs somewhat from our formulation. However, our definition is equivalent to the original one in \cite{Simon Stinchcombe:1995}, and closer to our definition of perfect equilibrium.}
A strategy profile $\sigma$ is called a \emph{weak perfect equilibrium} if there is a sequence $(\sigma^k)_{k=1}^\infty$ of strictly positive strategy profiles and a sequence $(\kappa^k)_{k=1}^\infty$ such that:
\vspace{-.15cm}
\begin{itemize}
\item[{[i]}] $\kappa^k$ is a best response to $\sigma^k$ for each $k \in \na$, and
\vspace{-.15cm}
\item[{[ii]}] both $d_P(\sigma^k,\sigma) \to 0$ and $d_P(\kappa^k,\sigma) \to 0$ as $k\to\infty$, where $d_P(\sigma,\tau)$ is the Prokhorov metric between $\sigma$ and $\tau$ for any two strategy profiles $\sigma, \tau \in \Delta$.
\vspace{-.15cm}
\end{itemize}

\begin{theorem}
\label{theo: weak perfect}
Let $\Gamma$ be a strategic form game with countably additive strategies as in Definition \ref{def-strformgamecad}. Let $\sigma$ be a perfect equilibrium of $\Gamma$ as in Definition \ref{def-perf}. Then $\sigma$ is a weak perfect equilibrium of $\Gamma$.
\end{theorem}

Proof. \quad
Suppose that $\sigma$ is a perfect equilibrium by Definition \ref{def-perf}. Let $\C$ be a carrier that includes the topology, and take $k \in \na$. Define
$
U_k = \Bigl\{ \tau \in \Delta \mid d_P(\tau, \sigma) < \frac{1}{k} \Bigr\}.
$
Then, by Theorem \ref{prop: characterization}, there are $\sigma^k, \kappa^k \in U_k$ such that $\C \subseteq \C(\sigma^k)$ and $\kappa^k$ is a best response to $\sigma^k$. Then $(\sigma^k)_{k=1}^\infty$ is a sequence of strictly positive strategies, and both $d_P(\sigma^k,\sigma) \to 0$ and $d_P(\tau^k,\sigma) \to 0$ as $k\to\infty$. Hence, $\sigma$ is weak perfect. 
\hfill \qed
\vskip6pt

{\eightrm TWO PLAYERS, ONE HAS ONLY TWO ACTIONS.} \quad
This section shows that our notion of perfection and that of weak perfect equilibrium \citep{Simon Stinchcombe:1995} coincide, when there are only two players, and one player has only two actions.\footnote{ It is an open question whether the two notions coincide for all games with countably additive strategies.}

\begin{theorem} \label{theo: two players}
Let $\Gamma = (N, A, \F, \Sigma, u)$ be a game with countably additive strategies as in Definition \ref{def-strformgamecad}, where $N = \{ 1, 2 \}$, and $A_1 = \{ a_1, b_1 \}$.
Let $\sigma = (\sigma_1, \sigma_2)$ be a strategy profile. Then $\sigma$ is a perfect equilibrium of $\Gamma$ as in Definition \ref{def-perf} precisely when it is a weak perfect equilibrium of $\Gamma$. In particular, strictly positive Nash equilibria are perfect equilibria.
\end{theorem}

Proof. \quad
By Theorem \ref{theo: weak perfect}, we know that any perfect equilibrium is a weak perfect equilibrium. We prove the reverse implication.
\vskip6pt

Since $\sigma$ is a weak perfect equilibrium, we can take a strictly positive sequence $(\sigma^k)_{k=1}^\infty$ and a sequence $(\eta^k)_{k=1}^\infty$ such that $\sigma^k \to \sigma$ and $\eta^k \to \sigma$ as $k \to \infty$, and $\eta^k$ is a best response to $\sigma^k$ for each $k$. Take any carrier $\C_2$ for player 2, and take any $\mu_2 \in \Delta_2$ with $\C_2(\mu_2) = \C_2$. 
\vskip6pt

We can assume, by taking a subsequence and changing the labels of actions $a_1$ and $b_1$ if necessary, that either both $a_1$ and $b_1$ are best responses to $\sigma_2^k$ for all $k$, or that $a_1$ is the unique best response to $\sigma_2^k$ for all $k$.
We discuss each case separately.
\vskip6pt

{\bf A.} \quad
Let $a_1$ be the unique best response against $\sigma_2^k$ for all $k$. For $k \in \na$, define $\kappa_2^k = \eta_2^k$, $\tau_1^k = \sigma_1^k$, and $\kappa_1^k = \delta(a_1)$. Define
$
\tau_2^k = (1-\veps_k) \cdot \sigma_2^k + \veps_k \cdot \mu_2,
$
with $\veps_k \in (0, \frac{1}{k})$ sufficiently small to guarantee that $a_1$ is still a best response to $\tau^k_2$. Fix any $k \in \na$. Then $\tau_1^k = \sigma_1^k$ has maximal carrier, and $\C(\tau_2^k)\supseteq \C(\mu_2) = \C_2$. Further, $\kappa_2^k = \eta_2^k$ is a best response to $\tau_1^k = \sigma_1^k$.
Also, by construction of $\tau_2^k$, we know that $\kappa_1^k = \delta(a_1)$ is a best response to $\tau_2^k$. It follows from item [3] of Theorem \ref{prop: characterization} that $\sigma$ is a perfect equilibrium according to Definition \ref{def-perf}.
\vskip6pt

The remaining case (B, see below) needs the next Claim.
\vskip6pt

{\bf Claim.} \quad
There is a strategy $\rho_2 \in \Delta_2$ with $\C_2(\mu_2) \subseteq \C_2(\rho_2)$ such that both $a_1$ and $b_1$ are best responses against $\rho_2$.
\vskip6pt

Proof of Claim. \quad
Assume wlog that $u_1(a_1, \tau_2) = 0$ for all $\tau_2 \in \Delta_2$.
Write $K = u_1(b_1, \mu_2)$. If $K = 0$, then take $\rho_2 = \mu_2$. If $K > 0$, then there must be a $b_2 \in A_2$ with $u_1(b_1, b_2) > 0$. Then, since $u_1$ is continuous, there is an open set $U_2 \ni b_2$ with $u_1(b_1, d_2) > 0$ for any $d_2 \in U_2$. Then, since $u_1(b_1, \sigma_2) = 0$, and $\sigma_2(U_2) > 0$, there must be a $c_2 \in A_2$ with $L = u_1(b_1, c_2) < 0$. Define
$
\rho_2 = \left[ \frac{-L}{K - L} \right] \cdot \mu_2 + \left[ \frac{K}{K - L} \right] \cdot \delta(c_2).
$
Note that $\rho_2$ is a strict convex combination of $\mu_2$ and $\delta(c_2)$. So, $\rho_2 \in \Delta_2$, and $\C_2(\mu_2) \subseteq \C_2(\sigma_2)$. Evidently $u_1(a_1, \rho_2) = 0$. Furthermore, since $K = u_1(b_1, \mu_2)$ and $L = u_1(b_1, c_2)$, we have that 
$
u_1(b_1, \rho_2) = 
\left[ \frac{-L}{K - L} \right] \cdot u_1(b_1, \mu_2) + \left[ \frac{K}{K - L} \right] \cdot u_1(b_1, c_2) = 0.
$
It follows that $a_1$ and $b_1$ are both best responses against $\rho_2$. A similar argument can be made in case that $K < 0$.
End proof of Claim.
\vskip6pt

{\bf B.} \quad
Let the pure best responses against $\sigma_2^k$ be $a_1$ and $b_1$, for all $k$. For $k \in \na$, $k \ge 1$, define $\kappa_1^k = \sigma_1$, $\kappa_2^k = \eta_2^k$, and $\tau_1^k = \sigma_1^k$. 
\vskip6pt

By the above claim, we can take a $\rho_2$ with $\C_2(\mu_2) \subseteq \C_2(\rho_2)$, and $a_1$ and $b_1$ are best responses against $\rho_2$. Define
$
\tau_2^k = \frac{k-1}{k} \cdot \sigma_2^k + \frac{1}{k} \cdot \rho_2.
$
Fix any $k \in \na$ with $k \ge 1$. Then $\tau_1^k = \sigma_1^k$ has maximal carrier, and $\C_2(\tau_2^k) \supseteq \C_2(\rho_2) \supseteq \C_2(\mu_2) = \C_2$. Further, $\kappa_2^k = \eta_2^k$ is a best response against $\tau_1^k = \sigma_1^k$.
Also, by construction of $\tau_2^k$, we know that both $a_1$ and $b_1$ are best responses against $\tau_2^k$. So, also $\kappa_1^k = \sigma$ is a best response against $\tau_2^k$. By item [3] of Theorem \ref{prop: characterization}, $\sigma$ is a perfect equilibrium according to Definition \ref{def-perf}.
\hfill \qed

\vspace{-.25cm}
\subsection{Games with finitely additive strategies}

For strategic form games with finitely additive strategies, this section relates our definition of perfect equilibrium to that of \cite{Marinacci: 1997}, referred to as M-perfection. The main result shows that every Nash equilibrium which is M-perfect
is also perfect according to Definition \ref{def-perf}. If each action set is at most countable, we show that any equilibrium in countably additive strategies is perfect precisely when the equilibrium is M-perfect.

\vspace{-.15cm}
\subsubsection*{The definition of Marinacci}

Let $\Gamma = (N, A, \F, \Xi, u)$ be a strategic form game with finitely additive strategies as in Definition \ref{def-strformgamefinad}. 
For $\veps_i > 0$, and $\tau_i \in \Xi_i$, the perturbed strategy space $\Xi_i(\tau_i, \veps_i)$ is given by
$
\Xi_i(\tau_i, \veps_i) =
\Bigl\{ (1 - \veps_i) \cdot \kappa_i + \veps_i \cdot \tau_i \mid \kappa_i \in \Xi_i \Bigr\}.
$
For a vector $\veps = (\veps_1, \ldots, \veps_n)$ we write $\veps > 0$ if $\veps_i > 0$ for all $i$.  For $\veps > 0$ and $\tau \in \Xi$ we write
$
\Xi(\tau, \veps) = \prod_{i \in N} \Xi_i(\tau_i, \veps_i).
$
We write $\Gamma(\tau, \veps) = (N, A, \F, \Xi(\tau, \veps), u)$ to denote the strategic form game in which each player $i$ is only allowed to choose strategies from within the set $\Xi_i(\tau_i, \veps_i)$.
\vskip6pt

Recall that a strategy $\tau_i \in \Xi_i$ has maximal carrier if $\C(\tau_i) = \F_i \setminus \{ \emptyset \}$.\footnote{In \cite{Marinacci: 1997} a strategy with maximal carrier is called strictly positive.}
A strategy profile $\tau = (\tau_1, \ldots, \tau_n)$ has maximal carrier if each $\tau_i$ has maximal carrier.
\vskip6pt

A Nash equilibrium $\kappa \in \Xi$ is perfect in the sense of Marinacci (M-perfect) if for every open $U \ni \kappa$ and every $\eta > 0$ there are a vector $\veps > 0$ with $\|\veps\| < \eta$ and a strategy profile $\tau \in \Xi$ such that the game $\Gamma(\tau, \veps)$ admits a Nash equilibrium $\sigma \in U$.\footnote{The original definition in \cite{Marinacci: 1997} does not require $\| \veps \| < \eta$. This however is no doubt a typo.}
\vskip6pt

{\bf Remark.} \quad
The definition requires that for any neighborhood of the equilibrium, there is a game with a restricted space of strategy profiles such that, for a sufficiently small restriction, the game has a Nash equilibrium in the specified neighborhood. 

\vspace{-.15cm}
\subsubsection*{The general result for games with finitely additive strategies}

\cite{Marinacci: 1997} showed the following result.
We say that a field $\F_i$ on $A_i$ is separable if there is a countable subset $D_i$ of $\F_i$ such that for every non-empty $F_i \in \F_i$ there is a non-empty $G_i \in D_i$ with $G_i \subseteq F_i$. 

\begin{theorem} \label{theo: Marinacci}
{\rm (Marinacci)} \quad
Let $\Gamma$ be a separable strategic form game with finitely additive strategies as in Definition \ref{def-strformgamefinad}. Then the set of M-perfect equilibria is non-empty and compact.
\end{theorem}

We prove the following statement (which does not require the separability condition).

\begin{theorem} \label{theo: M implies perfect}
Let $\Gamma$ be a strategic form game with finitely additive strategies as in Definition \ref{def-strformgamefinad}. Then every M-perfect equilibrium is a perfect equilibrium in the sense of Definition \ref{def-perf}.
\end{theorem}

Proof. \quad
Note that for any strategy $\tau_i$ with maximal carrier it holds that $\C_i(\tau_i) = \F_i \setminus \{ \emptyset \}$. The statement now follows directly from Theorem \ref{prop: characterization}. 
\hfill \qed
\vskip6pt

As \cite{Marinacci: 1997} observes, for strategic form games that are not separable, the set of M-perfect equilibria may well be empty. The next example illustrates this.

\begin{example} A game that does not have an M-perfect equilibrium. \rm
Consider the game from Example A in Section \ref{Section: Examples}, in which $\F_i = \B[0, 1]$ for $i = 1, 2$. So, for every $x \in [0, 1]$, the singleton set $\{ x \}$ is an element of $\F_i$. Hence, there is no $\tau \in \Xi$ that has maximal carrier and the set of M-perfect equilibria is empty.
\vskip3pt

This observation is independent of payoff functions. For \emph{any} strategic form game with $\F_i = \B[0, 1]$ for at least one player $i \in N$, the set of M-perfect equilibria is empty.
\hfill $\triangleleft$
\end{example}

\vspace{-.35cm}
\subsubsection*{A result for Nash equilibria in countably additive strategies}

Consider a strategic form game $\Gamma = (N, A, \F, \Xi, u)$ with finitely additive strategies, and $A_i = \na$ for all $i \in N$.
\vskip6pt

For a finitely additive measure $\zeta \ge 0$ on $\na$, we can write $\zeta = \zeta^c + \zeta^d$, where $\zeta^c$ is the countably additive part of $\zeta$, and $\zeta^d$ is the diffuse part of $\zeta$.\footnote{That is, $\zeta^c$ is a countably additive measure on $\na$, and $\zeta^d$ is a diffuse measure on $\na$. This decomposition is unique.}
A strategy $\kappa_i \in \Xi_i$ is countably additive if $\kappa_i^d = 0$. Equivalently, $\kappa_i \in \Delta_i$. A strategy profile is called countably additive of each strategy in that profile is countably additive. We show that such a strategy profile is a perfect equilibrium precisely when it is M-perfect.
\vskip6pt

To prove this, we need two preliminary results. We first show that perturbed games $\Gamma(\tau, \veps)$ can be represented by perturbed games $\Gamma(\zeta)$, where $\zeta = (\zeta_i, \ldots, \zeta_n)$ is a profile of finitely additive measures with $\zeta_i(\na) < 1$ for every $i \in N$. For each $i$, the measure $\zeta_i$ acts as a minimum weight on $\F_i$.
\vskip6pt

For a profile of finitely additive measures $\zeta = (\zeta_i, \ldots, \zeta_n)$ with $\zeta_i(\na) < 1$ for each $i$, write
$
\Xi_i(\zeta_i) =
\left\{ \eta_i \in \Xi_i \mid \eta_i(E) \ge \zeta_i(E) 
 \hbox{ for all } E \subseteq \na \right\}
$
for each $i$. We write $\Gamma(\zeta) = (N, A, \F, \Xi(\zeta), u)$ to denote the strategic form game in which each player $i$ is only allowed to choose strategies from within the set $\Xi_i(\zeta_i)$.

\begin{lemma} \label{lemm: mixing in equals restriction}
Let $\zeta = (\zeta_i, \ldots, \zeta_n)$ be a profile of finitely additive measures with $\zeta_i(\na) < 1$ for each $i$. Then $\Xi_i(\zeta_i) = \Xi_i(\phi_i, K_i)$, where $K_i= \zeta_i(\na)$ and $\phi_i= \frac{\zeta_i}{K_i}$. Consequently, for $K = (K_1, \ldots, K_n)$ it holds that $\Gamma(\phi, K) = \Gamma(\zeta)$.
\end{lemma}

Proof. \quad
Define the linear map $T_i \colon \Xi_i \rightarrow \Xi_i(\phi_i, K_i)$ by
$
T_i(\psi_i) = (1 - K_i) \cdot \psi_i + K_i \cdot \phi_i.
$
Then $T_i(\Xi_i) = \Xi(\phi_i, K_i)$ by definition of $\Xi_i(\phi_i, K_i)$. We show that $T_i$ is a bijective map from $\Xi_i$ to $\Xi_i(\zeta_i)$. Take any $\psi_i \in \Xi_i$. We show that $T_i(\psi_i) \in \Xi_i(\zeta_i)$. Note that for every $E \subseteq \na$ it holds that
$$
T_i(\psi_i)(E) = (1 - K_i) \cdot \psi_i(E) + K_i \cdot \phi_i(E) \ge K_i \cdot \phi_i(E) = \zeta_i(E). 
$$
So, $T_i(\psi_i) \in \Xi_i(\zeta_i)$. Conversely, take any $\rho_i \in \Xi_i(\zeta_i)$. Write $\lambda_i = \frac{K_i}{1 - K_i}$, and define $\psi_i = (1 + \lambda_i) \cdot \rho_i - \lambda_i \cdot \phi_i$. Then
$$
\psi_i(E) = 
(1 + \lambda_i) \cdot \rho_i(E) - \lambda_i \cdot \phi_i(E) = 
\frac{\rho_i(E)}{1 - K_i} - \frac{K_i \cdot \phi_i(E)}{1 - K_i} = 
\frac{\rho_i(E) - \zeta_i(E)}{1 - K_i} \ge 0. 
$$
Further, $\psi_i(\na) = 1 + \lambda_i - \lambda_i = 1$.
So, $\psi_i \in \Xi_i$. Further,
$
T_i(\psi_i) = (1 - K_i) \cdot \psi_i + K_i \cdot \phi_i = \rho_i + K_i \cdot \phi_i + K_i \cdot \phi_i = \rho_i.
$
This completes the proof.
\hfill \qed

\begin{lemma} \label{lemm: BR characterization}
Let $\rho \in \Xi(\zeta)$ be a strategy profile. Assume that for every player $i$, we have

\begin{itemize}
\vspace{-.15cm}
\item[{\rm [1]}] for every $k \in \na$ if $\rho_i^c(k) > \zeta_i^c(k)$, then $k$ is a best response to $\rho$, and
\vspace{-.15cm}
\item[{\rm [2]}] $\rho^d_i = \zeta_i^d$.
\vspace{-.15cm}
\end{itemize}
Then $\rho$ is a Nash equilibrium of the game $\Gamma(\zeta)$.
\end{lemma}

Proof. \quad
Let $\rho \in \Xi(\zeta)$ be a strategy profile for which [1] and [2] hold.\vskip6pt

Take any player $i$. We show that $\rho_i$ is a best response to $\rho$ within $\Xi_i(\zeta_i)$. Define $\delta_i = \rho_i - \zeta_i$. 
Then
$
\delta_i(\na) = \rho_i(\na) - \zeta_i(\na) = 1 - \zeta_i(\na) > 0.
$
Write $\kappa_i = \frac{\delta_i}{\delta_i(\na)}$. 
Since $\rho^d_i = \zeta_i^d$ by [2], $\delta_i = \rho_i^c - \zeta_i^c$.
It follows that $\kappa_i$ is countably additive. Take any $k$ with $\kappa_i(k) > 0$. We argue that $k$ is a best response to $\rho$ in $\Xi_i$. Since $\kappa_i(k) > 0$, it follows that $\rho_i^c(k) - \zeta_i^c(k) = \delta_i(k) > 0$. So, by [1], $k$ is a best response to $\rho$ in $\Xi$. Since $\kappa_i$ is countably additive, it follows that $\kappa_i$ is a best response to $\rho$ in $\Xi_i$. Then
$\rho_i = \delta_i(\na) \cdot \kappa_i + \zeta_i$ is a best response in $\Xi_i(\zeta_i)$ to $\rho$.
\hfill \qed
\vskip6pt

We can now prove the following theorem.

\begin{theorem} \label{theo: special Marinacci 1}
Let $\Gamma$ be a strategic form game in finitely additive strategies. Suppose that the action set $A_i$ of each player $i \in N$ is finite or countably infinite. Let $\sigma \in \Delta$ be a countably additive strategy profile. Then $\sigma$ is a perfect equilibrium of $\Gamma$ precisely when $\sigma$ is an M-perfect equilibrium of $\Gamma$.
\end{theorem}

Proof. \quad
Let $\sigma = (\sigma_1, \ldots, \sigma_n)$ be any countably additive perfect equilibrium. We assume without loss of generality that $A_i = \na$ for each $i \in N$.
\vskip6pt

Take an open set $U \ni \sigma$ and a real number $\eta > 0$. We show that there are a vector $K = (K_1, \ldots, K_n) > 0$ with $\|K\| < 3 \cdot \eta$ and a strategy profile $\rho \in \Xi$ with maximal carrier such that there is a Nash equilibrium $\rho \in \Gamma(\phi, K)$ with $\rho \in U$.
\vskip6pt

For the given $\eta > 0$, take $L \in \na$ such that
$
\sum_{k = L}^\infty \sigma_i(k) \le \eta
$
holds for each $i \in N$. Take $\delta \in (0, \frac{\eta}{L})$ such that, for all $i \in N$ and all $k < L$ with $\sigma_i(k) > 0$, it holds that
$
2 \cdot \delta < \sigma_i(k).
$
For each $i \in N$, let $U_i(L, \delta)$ be the open set of all strategies $\tau_i \in \Xi_i$ with
$
\left| \tau_i(k) - \sigma_i(k) \right| < \delta
$
for all $k < L$. Note that $\sigma_i \in U_i(L, \delta)$. We assume without loss of generality that $U = \prod_i U_i$ and that $U_i \subseteq U_i(L, \delta)$ for each $i \in N$.
\vskip6pt

Since $\sigma$ is perfect, we can take strategies $\rho_i \in U_i$ with maximal carrier and strategies $\kappa_i \in U_i$ such that
$\kappa$ is a best response to $\rho$.
\vskip6pt

For each $i$, write $\rho_i = \gamma_i + \xi_i$, where $\gamma_i$ is countably additive and $\xi_i$ is diffuse. Define $\zeta_i = \mu_i + \xi_i$, where
$$
\mu_i(k) =
\begin{cases}
\min \{ \gamma_i(k), \delta \} & \hbox{if } k < L \cr
\gamma_i(k) & \hbox{otherwise.}
\end{cases}
$$
{\bf Claim 1.} \quad
The strategy profile $\rho$ is a Nash equilibrium of $\Gamma(\zeta)$.
\vskip6pt

Proof of Claim 1. \quad
Since $\rho_i$ has maximal carrier, also $\gamma_i$ has maximal carrier. Then also $\mu_i$ has maximal carrier, and also $\zeta_i$ has maximal carrier. Second, note that $\gamma_i \ge \mu_i$, so that $\rho_i \in \Xi_i(\zeta_i)$.
\vskip6pt

We prove that $\rho_i$ is a best response against $\rho$ in $\Xi_i(\zeta_i)$. In order to do so, we check conditions [1] and [2] of Lemma \ref{lemm: BR characterization}.
\vskip6pt

In order to check [1], take any $k \in \na$ with $\rho_i(k) > \zeta_i(k)$. Then $\gamma_i(k) > \mu_i(k)$. Then $k < L$ by definition of $\mu_i$, and $\mu_i(k) = \delta$. Suppose that $k$ is not a best response to $\rho$. As $\kappa$ is a best response to $\rho$, we know that $\kappa_i(k) = 0$. 
Then, since $\kappa_i \in U_i$, it follows that $\sigma_i(k) < \delta$. Then, since $k < L$, and
$
2 \cdot \delta < \sigma_i(k) \quad
\hbox{for all } k
$
with $\sigma_i(k) > 0$, it follows that $\sigma_i(k) = 0$. Then, since
$
\rho_i(k) > \zeta_i(k) = \mu_i(k) = \delta,
$
we conclude that $\rho_i \notin U_i$. This is a contradiction. So, [1] of Lemma \ref{lemm: BR characterization} holds. Next, note that $\rho_i^d = \xi_i = \zeta^d_i$. So, [2] of Lemma \ref{lemm: BR characterization} holds as well. End proof of Claim 1.
\vskip6pt

So, we know that $\rho$ is a Nash equilibrium of $\Gamma(\zeta)$. 
Define $K_i= \zeta_i(\na)$ and $\phi_i= \frac{\zeta_i}{K_i}$. Then by Lemma \ref{lemm: mixing in equals restriction} we know that $\rho$ is a Nash equilibrium of $\Gamma(\phi, K)$.
\vskip6pt

{\bf Claim 2.} \quad
It holds that $\| K \| < 3 \cdot \eta$.
\vskip6pt

Proof of Claim 2. \quad
Take any player $i$. Recall that
$
\sum_{k = L}^\infty \sigma_i(k) \le \eta
$
and that $\delta \in (0, \frac{\eta}{L})$. Using $\rho_i \in U_i$ and the fact that $\sigma_i$ is countably additive,
$$
\sum_{k=1}^{L-1} \rho_i(k) \ge \sum_{k=1}^{L-1} \sigma_i(k) - (L-1) \cdot \delta > \sum_{k=1}^{L-1} \sigma_i(k) - \eta \ge 1 - 2 \cdot \eta.
$$
It follows that
$
\xi_i(\na) + \sum_{k=L}^\infty \rho_i(k) = 1 - \sum_{k=1}^{L-1} \rho_i(k) < 2 \cdot \eta.
$
Hence,
$$
K_i =
\zeta_i(\na) = \mu_i(\na) + \xi_i(\na) \le
\sum_{k=1}^{L_i - 1} \mu_i(k) + \sum_{k=L}^\infty \rho_i(k) + \xi_i(\na)
< \delta \cdot L + 2 \cdot \eta \le 3 \cdot \eta.
$$
End proof of Claim 2.
\hfill \qed


\section{Summary and conclusion}\label{Section: Conclusion}
This paper proposed a new definition of perfect equilibrium. Following Selten's (1975) original idea, it selects Nash equilibria that are robust to opponents making mistakes in their choice of equilibrium actions. A key feature of our definition is that a perfect equilibrium is robust to all kinds of such mistakes, including for instance deviations taking the form of a Dirac measure.
\vskip6pt

What enables this robustness is our new interpretation of what defines the carrier of a strategy. Rather than defining it in the usual way, we define it as the set of all measurable sets to which a strategy assigns strictly positive probability. This allows for a general proof of existence of perfect equilibrium, owing to a feature related to the finite intersection property. The definition of carrier furthermore warrants a clear connection between perfection and pure game theoretic notions such a limit undominatedness and invariance.
\vskip6pt

For games with finite actions sets, our definition of perfect equilibrium is equivalent to Selten's (1975) original definition and several alternative definitions from the literature. For games with infinite action sets the relation to earlier definitions is less clear. In the classic setting of games with countably additive strategies, every equilibrium which is perfect in our sense is weakly perfect in the sense of \cite{Simon Stinchcombe:1995}. The two notions even coincide for games where there are only two players and at least one player only has two actions. But, whether or not this is true in general remains an open question. Another open question is whether for games which finitely additive strategies and countable action sets, our definition of perfect equilibrium is equivalent to Marinacci's (1997) definition.
\vskip6pt

Yet further open questions arise from possible extensions of the definition of perfect equilibrium to more general classes of games, such as signaling games or extensive form games.


\clearpage

\begin{center}
{\bf\Huge Supplemental Appendix}
\end{center}
\section*{\large Appendix A. \quad Detailed Proofs for Section \ref{Section: Standard frameworks}, finite additivity}

Consider a game $\Gamma^\ast = (N, A, \F, \Sigma, u)$, where $N$, $A$, $\F$, and $\Sigma$ are as in Definition \ref{def-strformgame}. Further, $u = (u_1, \ldots, u_n)$ is a profile of functions where for each player $i\in N$, $u_i \colon A \rightarrow \re$ is a bounded payoff function. For completeness, we first provide a detailed definition of viability of $u_i$.
For each simple function $f_i = \sum_{\ell=1}^k c_\ell \cdot \id_{R^\ell}$ and strategy profile $\kappa = (\kappa_i)_{i \in N} \in \Sigma$, we define
\begin{equation}\label{def-simpleexp}
\kappa(f_i) := \sum_{\ell=1}^k c_\ell \cdot \kappa(R^\ell),
\quad \hbox{where }
\kappa(R^\ell) := \prod_{i \in N} \kappa_i(R_i^\ell).
\end{equation}
We define
\begin{eqnarray*}
\alpha_i(\kappa)
& := &
\sup \left\{ \kappa(f_i) \mid f_i\text{ is simple}, f_i \le u_i \right\} \cr
\hbox{ and } \quad \quad
\beta_i(\kappa)
& := &
\ \inf \left\{ \kappa(g_i) \mid g_i\text{ is simple}, g_i \ge u_i \right\}.
\end{eqnarray*}
Note that $\alpha_i(\kappa)$ and $\beta_i(\kappa)$ are well defined as $u_i$ is bounded. Clearly, $\alpha_i(\kappa) \le \beta_i(\kappa)$.
The real number $\alpha_i(\kappa)$ is the lower integral of $u_i$ over $A$ with respect to $\kappa$, and $\beta_i(\kappa)$ is the upper integral of $u_i$ over $A$ with respect to $\kappa$. If $\alpha_i(\kappa) = \beta_i(\kappa)$, then $u_i$ is $\kappa$-integrable.
\vskip6pt

Thus, $u_i$ is viable precisely when $u_i$ is $\kappa$-integrable for each $\kappa \in \Sigma$.
For a viable payoff function $u_i$ we define the expected payoff function $U_i$, for all $\kappa \in \Sigma$, via
$$
U_i(\kappa) := \alpha_i(\kappa) = \beta_i(\kappa).
$$
Clearly, $U_i(\kappa)$ is indeed the integral of $u_i$ with respect to $\kappa$. It is well-known that, if $u_i$ is a viable payoff function for some player $i\in N$, then $U_i$ is multi-linear. A short proof of this fact is provided in the Lemma below.

\begin{lemma} \label{viable is multilinear}
Let $u_i$ be a viable payoff function for some player $i\in N$. Then, $U_i$ is multi-linear.
\end{lemma}

\noindent
Proof. \quad
Take a strategy profile $\kappa \in \Sigma$, a strategy $\mu_j \in \Sigma_j$ for some player $j\in N$, some $\lambda \in [0,1]$, and let $\zeta_j := \lambda \cdot \kappa_j + (1 - \lambda) \cdot \mu_j$. We need to show that
\begin{equation}\label{eq-toprove}
U_i(\zeta_j \mid \kappa) \,=\,
\lambda \cdot U_i(\kappa) + (1 - \lambda) \cdot U_i(\mu_j\mid \kappa).
\end{equation}

Let $\delta>0$. Since the maximum of two simple functions is also a simple function, by the definition of $\alpha_i$, there is a simple function $f_i$ such that (i) $f_i\leq u_i$, (ii) $\kappa(f_i)\geq \alpha_i(\kappa)-\delta$, and (iii)  $(\mu_j\mid \kappa )(f_i)\geq \alpha_i(\mu_j\mid \kappa )-\delta$.  Hence, 
\begin{align*}
U_i(\zeta_j\mid \kappa )\,&=\,\alpha_i(\zeta_j\mid \kappa )\,\geq\,( \zeta_j\mid \kappa )(f_i)\,=\,\lambda \cdot \kappa(f_i) + (1 - \lambda) \cdot (\mu_j\mid \kappa)(f_i)\\[0.1cm]
&\geq\,\lambda \cdot \alpha_i(\kappa)+(1 - \lambda) \cdot \alpha_i(\mu_j\mid \kappa )-\delta
=\,\lambda \cdot U_i(\kappa)+(1 - \lambda) \cdot U_i(\mu_j\mid \kappa )-\delta,
\end{align*}
where the first equality is by viability of $u_i$, the first inequality is by definition of $\alpha_i$ and (i), the second equality is by \Eqref{def-simpleexp}, the second inequality is by (ii) and (iii), and the last equality is by viability of $u_i$ again. Using $\beta_i(\cdot)$ instead of $\alpha_i(\cdot)$, one also finds $U_i(\zeta_j\mid \kappa )\leq \lambda \cdot U_i(\kappa)+(1 - \lambda) \cdot U_i(\mu_j\mid \kappa )+\delta$. As $\delta>0$ was arbitrary, we have proven \Eqref{eq-toprove}, as desired.
\hfill \qed
\vskip6pt

Next, we provide a proof of the following well-known Lemma.

\begin{lemma}
\label{theo: existence for finite additivity}
If $u_i$ is a uniform limit of simple functions, then $u_i$ is viable, and the expected payoff function $U_i$ is multi-linear, and continuous in the Tychonov topology.
\end{lemma}

Proof. \quad
We first show that $u_i$ is viable, so that $U_i$ exists.
Take any $\veps > 0$. Since $u_i$ is a uniform limit of simple functions, there is a simple function $f_i$ with
$$
f_i(a) - \veps \le u_i(a) \le f_i(a) + \veps
\quad\quad \hbox{ for every } a \in A.
$$
Take any $\kappa \in \Xi$. It follows that
$
\alpha_i(\kappa) \ge \kappa(f_i - \veps \cdot \id_A) = \kappa(f_i) - \veps.
$
In the same way we find that $\beta_i(\kappa) \le \kappa(f_i) + \veps$.
It follows that
$
\beta_i(\kappa) - \alpha_i(\kappa) \le 2 \cdot \veps.
$
Since $\veps > 0$ was chosen arbitrarily and $\beta_i(\kappa) \ge \alpha_i(\kappa)$, it follows that $\beta_i(\kappa) = \alpha_i(\kappa)$. Thus, $u_i$ is viable.
So, the expected payoff function $U_i$ exists, and is hence multi-linear.
\vskip6pt

We prove that $U_i$ is continuous. First note that every simple function $f_i \colon \Sigma \rightarrow \re$ is continuous. Then, since $U_i$ is a uniform limit of continuous functions, also $U_i$ is continuous.
\hfill \qed

\subsection*{Detailed Proofs for Example \ref{example: variant Wald game}}

Proof of Claim 1. \quad
The function $u_1$ is the uniform limit of the simple functions $u_1^n$ defined by
$$
u_1^n(k, \ell) =
\begin{cases}
u_1(k, \ell) & \hbox{if } k \le n \hbox{ and } \ell \le n \cr
0 & \hbox{otherwise.}
\end{cases}
$$
End proof of Claim 1.
\hfill \qed
\vskip6pt

Proof of Claim 2. \quad
Let $(\kappa_1, \kappa_2)$ be any strategy pair.
Suppose the opposite, that for each player $i$ there is a $k_i$ with $\kappa_i(k_i) > 0$. Let $\ell_i$ be the smallest natural number on which $\kappa_i$ puts strictly positive weight. Suppose wlog that $\ell_1 \le \ell_2$. Then player 1 can strictly improve his payoff by moving the probability $\kappa_1(\ell_1)$ from $\ell_1$ to $\ell_2 + 1$. Hence,
$(\kappa_1, \kappa_2) \in \Xi$ is not a Nash equilibrium. 
End proof of Claim 2.
\hfill \qed
\vskip6pt

Proof of Claim 3. \quad
Since one player playing a diffuse charge results in payoff $0$ for both players, it is easy to check that any pair of diffuse charges is a Nash equilibrium.
\vskip6pt

Conversely, let $\kappa = (\kappa_1, \kappa_2)$ be a Nash equilibrium. From Claim 2 we know that at least one player plays a diffuse charge, say $\kappa_1$ is a diffuse charge. Suppose that $\kappa_2$ is not a diffuse charge, say $\kappa_2(k) > 0$. Then the payoff for player 1 playing $\delta(k+1)$ is strictly positive, and hence strictly higher than the payoff for $\kappa_1$. So, $\kappa_2$ must be a diffuse charge as well.
End proof of Claim 3.
\hfill \qed
\vskip6pt

Proof of Claim 4. \quad
By Claim 3 it suffices to prove that every pair of diffuse charges is a perfect strategy pair.
Let $\mu = (\mu_1, \mu_2)$ be a pair of diffuse charges. We argue that $\mu$ is a perfect strategy pair.
\vskip6pt

We prove [2] of Theorem \ref{prop: characterization}. Let a carrier $\C_1$ for player 1 and open sets $U_1 \ni \mu_1$ and $U_2 \ni \mu_2$ be given. We construct strategies $\tau_1 \in U_1$ and $\kappa_2 \in U_2$ such that $\C_1 \subseteq \C(\tau_1)$ and $\kappa_2$ is a best response against $\tau_1$. Part [2] of Theorem \ref{prop: characterization} then follows by symmetry.
\vskip6pt

We can take a partition $V_1, \ldots, V_M$ of $\na$ into $M$ sets and $\veps > 0$ such that
$$
W_2 := \Bigl\{ \kappa_2 \ \mid \ \Bigl| \kappa_2(V_s) - \mu_2(V_s) \Bigr| < \veps \hbox{ for all } s = 1, \ldots, M \Bigr\} \subseteq U_2.
$$
For each $s$, define $q_s = \mu_2(V_s)$. Take $K_s \in V_s$. We assume with loss of generality that $K_1 < K_2 < \cdots < K_M$. Take $\sigma_1 \in \Delta(\na)$ as in Lemma \ref{lemm: technical} below. Define
$
\kappa_2 = \sum_{s=1}^M q_s \cdot \delta(K_s).
$
Then clearly $\kappa_2 \in W_2$, and $\kappa_2$ is a best response against $\sigma_1$. Choose $\eta \in (0, 1)$ sufficiently small such that
$
\tau_1 := \eta \cdot \sigma_1 + (1-\eta) \cdot \mu_1 \in U_1.
$
Note that, since $\mu_1$ is diffuse, any strategy of player 2 yields payoff zero against $\mu_1$. As such, $\kappa_2$ is a best response to $\mu_1$. Then, since the expected payoff function of player 2 is linear in the strategy of player 1, $\kappa_2$ is a best response to $\tau_1$. Moreover, since $\C(\sigma_1) = 2^{\na} \setminus \{ \emptyset \}$, it holds that $\C_1 \subseteq 2^{\na} \setminus \{ \emptyset \} = \C(\tau_1)$.
End proof of Claim 4.
\hfill \qed

\begin{lemma} \label{lemm: technical}
Let $K_1, \ldots, K_M$ be natural numbers with $1 < K_1 < K_2 < \cdots < K_M$. Then there is a $\sigma_i \in \Delta(\na)$ with maximal carrier against which the pure best responses are precisely the points $K_1, \ldots, K_M$.
\end{lemma}

Proof. \quad
Note that a strategy $\sigma_i$ has maximal carrier precisely when $\sigma_i(k) > 0$ for all $k \in \na$. In the proof we use the following notation.
For $\sigma_i \in \Delta(\na)$ and for each $k \ge 2$, write $P(k) = \sum_{m=1}^{k-1} \sigma_i(m)$. Then the payoff to player $j$ of playing $k$ is $\frac{P(k)}{k}$. The proof is in two parts.
\vskip6pt

{\bf A.} \quad 
Suppose that $\sigma_i(k) = \sigma_i(k+1) = \cdots\cdots = \sigma_i(k+m) = z$. Then $k+m$ is a better response than $k+m+1$ precisely when $P(k) < k \cdot z$.
\vskip6pt

Proof of claim. Strategy $k+m$ is a better response than $k+m+1$ precisely when 
$$
\frac{P(k)+m \cdot z}{k+m} < \frac{P(k)+(m+1) \cdot z}{k+m + 1}.
$$
This can be rewritten to $P(k) < k \cdot z$. End of proof claim.
\vskip6pt

{\bf B.} \quad
We define $\sigma_i$ by induction. Recall that $K_1 > 1$. For all $k = 1, \ldots, K_1-1$, we define $\sigma_i(k) = \eta$. The payoff to player $j$ of playing $k$ is $\frac{P(k)}{k} = \frac{k-1}{k} \cdot \eta$. So, $K_1$ is the best response among the strategies $k = 1, \ldots, K_1$.
\vskip6pt

Suppose that $\sigma_i(k)$ is defined for all $k = 1, \ldots, K_m -1$, and that the best responses of player $j$ to $\sigma_i$ are $K_1, \ldots, K_m$. We define $\sigma_i(k)$ for $k = K_m, \ldots, K_{m+1}-1$ in such a way that the expected payoff to player $j$ for $K_{m+1} = K_m + \ell$ is equal to the expected payoff for $K_m$, and strictly less for all intermediate $k = K_m +1, \ldots, K_{m+1} - 1$. Take
$$
\sigma_i(K_m) = \cdots = \sigma_i(K_m+\ell -2) = \tau
\quad \hbox{and} \quad
\sigma_i(K_m+\ell-1) =
(\ell+1) \cdot \tau,
$$
where
$
\tau = \frac{P(K_m)}{2 K_m}.
$
Then 
\begin{eqnarray*}
P(K_{m+1})
& = &
P(K_m) + (\ell-1) \cdot \tau + (\ell+1) \cdot \tau = P(K_m) + 2 \ell \tau \cr
&& \cr
& = &
P(K_m) \cdot \left( 1 + \frac{\ell}{K_m} \right) = 
P(K_m) \cdot \frac{K_m+\ell}{K_m} = P(K_m) \cdot \frac{K_{m+1}}{K_m}.
\end{eqnarray*}
So, 
$$
\frac{P(K_{m+1})}{K_{m+1}} = \frac{P(K_m)}{K_m},
$$
which means that the payoff to player $j$ for strategy $K_{m+1} = K_m + \ell$ equals the payoff to player $j$ of strategy $K_m$. 
\vskip6pt

The payoff to player $j$ for a strategy $k$ with $K_m < k < K_m+\ell$ is
$$
\frac{P(K_m)}{k} + \frac{k - K_m - 1}{k} \cdot \frac{P(K_m)}{2K_m} < \frac{P(K_m)}{K_m} \cdot \left( \frac{K_m}{k} + \frac{k - K_m}{k} \right) = \frac{P(K_m)}{K_m}.
$$
So, $k$ has a strictly lower payoff for player $j$ than strategy $K_m$. 
\vskip6pt

Since we only defined $\sigma_i$ so far for strategies $k = 1, \ldots, K_M - 1$, we can define
$$
\sigma_i(k) = \frac{P(K_M)}{2 K_M} \cdot \left( \frac{1}{2} \right)^k \cdot \eta \quad \hbox{ for all } k \ge K_M
$$
and then choose $\eta$ such that $\sum_{k=1}^\infty \sigma_i(k) = 1$.
\hfill \qed

\vspace{-.25cm}
\section*{\large Appendix B. \quad Detailed Proofs for Section \ref{Section: Standard frameworks}, countable additivity}

Let $B$ be a subset of $A$. A point $b \in A$ is called a boundary point
of $B$ if for every $\veps > 0$ there are $x \in B$ and $y \in A \setminus B$ with
$d(b, x) < \veps$ and $d(b, y) < \veps$. The set of boundary points of $B$ is denoted by $\partial B$.
Let $\sigma$ be a strategy profile. A set $R \subseteq A$ is called a continuity set for $\sigma$ if $\sigma(\partial R) = 0$.

\begin{lemma} \label{lemm: regularity}
Let $B$ and $D$ be two sets that are continuity sets for $\sigma$.
Then also $A \setminus B$ and $B \cap D$ are continuity sets for $\sigma$.
\end{lemma}

\noindent
Proof. \quad
The continuity of $A \setminus B$ is evident from the definitions. We show that $E = B \cap D$ is a continuity set for $\sigma$. Let $b \in \partial E$. Take sequences $x^n \in E$ and $y^n \notin E$ with $d(x^n, b) < \frac{1}{n}$ and $d(y^n, b) < \frac{1}{n}$.
Since $y^n \notin E$ for all $n$, by taking a subsequence we can assume that $y^n \notin B$ for all $n$, or $y^n \notin D$ for all $n$.
\vskip6pt

\noindent
If $y^n \notin B$ for all $n$, then, since $x^n \in B$ for all $n$, while
$d(x^n, b) < \frac{1}{n}$ and $d(y^n, b) < \frac{1}{n}$, it follows that $b \in \partial B$. If $y^n \notin D$ for all $n$, we conclude in the same way that $b \in \partial D$. Thus, $\partial E$ is a subset of $\partial B \cup \partial D$, which has probability zero according to $\sigma$.
\hfill \qed

\begin{lemma} \label{lemm: partition}
Let $B^1, \ldots B^\ell$ be a collection of rectangles in $A$. Then there are collections $R^1, \ldots, R^M$ and $T^1, \ldots, T^L$ of non-empty rectangles such that
\begin{itemize}
\vspace{-.15cm}
    \item[{\rm [1]}] the collection $R^1, \ldots, R^M, T^1, \ldots, T^L$ is a partition of $A$
\vspace{-.15cm}
    \item[{\rm [2]}] for each $R^m$ there is at least one $B^k$ with $R^m \subseteq B^k$
\vspace{-.15cm}
    \item[{\rm [3]}] for each $T^l$ and $B^k$, $T^l \cap B^k = \emptyset$
\vspace{-.15cm}
    \item[{\rm [4]}] for each $R^m$ and $B^k$, either $R^m \subseteq B^k$ or $R^m \cap B^k = \emptyset$.
\vspace{-.15cm}
\end{itemize}
If each $B^k$ is regular for $\sigma$, then so are each $R^m$ and $T^l$.
\end{lemma}
\vskip6pt

\noindent
Proof. \quad
Let $B^1, \ldots B^\ell$ be given. For each $i \in N$, take a set $S(i) \subseteq \{ 1, \ldots, \ell \}$. Write $S = (S(1), \ldots, S(n)) \in {\mathcal S}(\ell)$. Define
$
R(S) = \prod_{i \in N} \left[ \bigcap_{k \in S(i)} B^k_i \cap
\bigcap_{k \notin S(i)} \left( A \setminus B^k_i \right) \right].
$
Many of the sets $R(S)$ thus constructed may be empty. However, the non-empty sets $R(S)$ partition $A$, and all these sets are rectangles. Moreover, if each $B^k$ is regular for $\sigma$, then by Lemma \ref{lemm: regularity} so is each $R(S)$. We can divide the remaining non-empty rectangles $R(S)$ into those that are contained in some $B^k$, when $k \in S(i)$ for all $i$, and those that are not contained in any $B^k$, when $k \notin S(i)$ for at least one $i$. Observe that the resulting collections have the properties that are stated above. 
\hfill \qed
\vskip6pt

\noindent
The next Lemma is in fact a direct consequence of the Portmanteau Theorem for continuity sets. For completeness we provide a (short) direct proof.

\begin{lemma} \label{lemm: approximate regular}
Let $\sigma$ be a strategy profile and $R$ a rectangle that is regular for $\sigma$. Then for every $\veps > 0$ there is $\delta > 0$ such that:\;
$
\left| \sigma(R) - \tau(R) \right| < 2 \cdot \veps
$
whenever $d_P(\sigma, \tau) < \delta$.
\end{lemma}

\noindent
Proof. \quad
Let $\veps > 0$. Since $R$ is regular for $\sigma$, we know that
$$
\sigma(R_\delta) \rightarrow \sigma(R)
\quad \hbox{as} \quad \delta \downarrow 0.
$$
So, we can take $\delta < \veps$ such that $\sigma(R_\delta) < \sigma(R) + \veps$. Take any $\tau$ with $d_P(\sigma, \tau) < \delta$. Then
$
\tau(R) \le \sigma(R_\delta) + \delta < \sigma(R) + \veps + \veps.
$
In the same way we can show that $\tau(A \setminus R) < \sigma(A \setminus R) + 2 \cdot \veps$. It follows that $\tau(R) > \sigma(R) - 2 \cdot \veps$.
\hfill \qed
\vskip6pt

Next we prove the following well-known Lemma.
\vskip6pt

\begin{lemma} \label{theo: alpha is beta}
If $u_i$ is continuous, then $u_i$ is viable. Moreover, the induced expected payoff function $U_i \colon \Delta \rightarrow \re$ is multi-linear, and continuous in the Prohorov topology.
\end{lemma}

Proof. \quad
Let $u_i$ be continuous. Take a fixed $\sigma \in \Delta$.
We first show that $\alpha_i(\sigma) = \beta_i(\sigma)$.
\vskip6pt

\noindent
The proof is in several steps. Take any $a \in A$, and any $\veps > 0$.
Since $\sigma(A) = 1$, we know that there is a $M > 0$ such that $\sigma(A \setminus D_M(a)) < \veps$. Write $D = D_M(a)$ and $D_i = D_M(a_i)$. Then $D = \prod_i D_i$. Moreover, $D_i \in \F_i$, so that $D$ is a rectangle, and, since $A_i$ is locally compact, each $D_i$ is compact.
\vskip6pt

\noindent
Since $D$ is compact, we know that $u_i$ is uniformly continuous on $D$. 
So, there is a $\delta > 0$ such that for all $b, c \in D$ it holds that
$
| u_i(b) - u_i(c) | < \veps
\quad \hbox{ whenever } \quad
| b - c | < 2 \cdot \delta.
$
Then the collection of sets
$
D_{\delta}(b) = \{ d \in A \mid d(b, a) < \delta \}
$
with $b \in D$ covers the compact set $D$. So, we can select a finite number of points $b^1, \ldots, b^\ell$ such that the sets $D^k = D_{\delta}(b^k) \cap D$ still cover $D$.
\vskip6pt

Note that each $D^k$ is a rectangle. Then, by Lemma \ref{lemm: partition} 
there is a collection $R^1, \ldots, R^M$ of non-empty rectangles each of which is contained in some $D^k$, and a collection $T^1, \ldots, T^L$ of rectangles each of which has empty intersection with any $D^k$, such that these collections taken together form a partition of $A$.
For each $R^\ell$, take a point $r^\ell \in R^\ell$. Note that 
there is a $D^k$ that contains $R^\ell$. Thus, for any $c \in R^\ell$ it holds that
$
d(c, r^\ell) \le d(c, b^k) + d(b^k, r^\ell) < \delta + \delta = 2 \cdot \delta.
$
Define $c_k = u_i(r^k)$. Further, take $K > 0$ such that $- K \le u_i(a) \le K$ for all $a \in A$. Define
$
g_i = \sum_{k=1}^M c_k \cdot \id_{R^k}
\quad \hbox{and} \quad
h_i = g_i - \veps \cdot \id_D - \sum_{l = 1}^L K \cdot \id_{T^l}.
$
Then $g_i$ and $h_i$ are simple functions. Moreover, since $D$ and the sets $D^k$ can be assumed to be regular for $\sigma$, also all sets $R^m$ and $T^l$ may be assumed to be regular for $\sigma$.
\vskip6pt

\noindent
{\bf A.} \quad
We claim that
$
\alpha_i(\sigma) \ge \sigma(g_i) - (K+1) \cdot \veps
\quad \hbox{ and } \quad
\beta_i(\sigma) \le \sigma(g_i) + (K+1) \cdot \veps.
$
\vspace{.15cm}

Proof of claim A. \quad
Take any point $b \in D$. There is precisely one $k$ with $b \in R^k$. Hence, since $d(b, r^k) < \delta$, we have that
$
u_i(b) \ge u_i(r^k) - \veps = c_k - \veps = h_i(b).
$
It follows that $h_i$ is a lower bound of $u_i$. Since the union over all sets $T^l$ equals $A \setminus D$, we have
\begin{eqnarray*}
\alpha_i(\sigma) \ge \sigma(h_i)
=
\sigma( h_i \cdot \id_D) + \sigma( h_i \cdot \id_{A \setminus D} ) 
\ge
\sigma(g_i) - \veps - K \cdot \sigma(A \setminus D) 
=
\sigma(g_i) - \veps - K \cdot \veps.
\end{eqnarray*}
Similarly, we find that $\beta_i(\sigma) \le \sigma(g_i) + \veps + K \cdot \veps$. This ends the proof of claim A.
\vskip3pt

\noindent
{\bf B.} \quad
From claim A it follows that
$
\beta_i(\sigma) - \alpha_i(\sigma) \le 2 (K+1) \cdot \veps.
$
Since $\veps > 0$ was arbitrary, we have that $\beta_i(\sigma) \le \alpha_i(\sigma)$. Hence, the payoff function $u_i$ is viable.
\vskip3pt

\noindent
{\bf C.} \quad
We prove that the resulting expected payoff function $U_i$ is continuous.
Take $g_i$ and $h_i$ as before. We now assume that all sets $R^k$ and $T^l$ are regular. Then by Lemma \ref{lemm: approximate regular} we can take $\delta_1 > 0$ such that
$
|\sigma(R^k) - \tau(R^k) | < 2 \cdot \veps
$
for every $k = 1, \ldots, \ell$ whenever $d_P(\sigma, \tau) < \delta_1$.
We can also assume that the same holds for the set $A \setminus D$.
Take any $\tau$ with $d_P(\sigma, \tau) < \delta_1$. Note that
$
\tau(h_i) = \tau(g_i) - \veps \cdot \tau(D) - K \cdot \tau(A \setminus D).
$
Then, using claim A and the fact that $\alpha_i(\sigma) = \beta_i(\sigma)$, we have that
\begin{eqnarray*}
\alpha_i(\sigma) \le \sigma(g_i) + (K+1) \cdot \veps
& \le &
\tau(g_i) + (K+3) \cdot \veps \cr
& \le &
\tau(h_i) + \veps + K \cdot \tau(A \setminus D) + (K+3) \cdot \veps \cr
& \le &
\alpha_i(\tau) + K \cdot (\sigma(A \setminus D) + 2 \cdot \veps) + (K+4) \cdot \veps \cr
& \le &
\alpha_i(\tau) + K \cdot 3 \cdot \veps + (K+4) \cdot \veps 
=
\alpha_i(\tau) + 4 \cdot (K+1) \cdot \veps
\end{eqnarray*}
In the same way, we can find $\delta_2 > 0$ such that
for any $\tau$ with $d_P(\sigma, \tau) < \delta_2$ it holds that
$
\alpha_i(\sigma) \ge \alpha_i(\tau) - 4 \cdot (K+1) \cdot \veps
$
It follows for any $\tau$ with $d_P(\sigma, \tau) < \min\{ \delta_1, \delta_2 \}$ that
$
\left| \alpha_i(\sigma) - \alpha_i(\tau) \right| < 4 \cdot (K+1) \cdot \veps.
$
This completes the proof.
\hfill \qed

\vspace{-.25cm}
\section*{\large Appendix C. \quad A Fubini result}

This appendix proves the Fubini result in the context of finitely additive probability measures. Suppose that $N$ is partitioned into two non-empty sets $I$ and $J$. For a function $u_i \colon A \rightarrow \re$, define $u_i(b_I) \colon A_J \rightarrow \re$ by
$
u_i(b_I)(b_J) = u_i(b_I, b_J)
\; \hbox{for all } b_J \in A_J.
$

\subsection*{Simple functions}

A simple function $f_i$ is of product type when for each $i$ there is a partition $R_i^1, \ldots, R_i^{\ell_i}$ of $A_i$ by elements of $\F_i$ such that
$
f_i = \sum_k c_k \cdot \id_{R^k}
$
where $k =(k_1, \ldots, k_n)$ and 
$
R^k = \prod_{i=1}^n R_i^{k_i}.
$
It is straightforward to show that each simple function can be written in this form. In other words, each simple function has a product type representation.

\begin{lemma} \label{lemm: fibi}
Let $f_i$ be a simple function. Let $b_I \in A_I$. Then $f_i(b_I)$ is a simple function. Further, for each $\kappa_J \in \Xi_J$ it holds that:\;
$
\kappa_J(f_i(b_I)) = (\delta(b_I), \kappa_J)(f_i).
$
\end{lemma}

Proof. \quad
Let $b_I \in A_I$ and $\kappa_J \in \Xi_J$. Let
$
f_i = \sum_k c_k \cdot \id_{R^k}
$
be a product type representation of $f_i$. For each $i \in I$, take $t_i$ with $b_i \in R_i^{t_i}$. Then
$
f_i(b_I)(b_J) = \sum_{k_J} c_{t_I, k_J} \cdot \id_{R^{k_J}}
$
is a simple function in product type representation. Further, write
$
\kappa_J(R_J^{k_J}) = \prod_{j \in J} \kappa_j(R_j^{k_j}).
$
Then
\begin{eqnarray*}
f_i(b_I)(\kappa_J)
=
\sum_{k_J} c_{t_I, k_J} \cdot \kappa_J(R_J^{k_J}) =
\sum_{k_J} c_{t_I, k_J} \cdot \kappa_J(R_J^{k_J}) \cdot \prod_{i \in I} \delta(b_i)(R_i^{t_i})
=
(\delta(b_I), \kappa_J)(f_i).
\end{eqnarray*}
This concludes the proof.
\hfill \qed
\vskip6pt

Let $\kappa_J \in \Xi_J$. Define $g_i(\kappa_J) \colon A_I \rightarrow \re$ by
$
g_i(\kappa_J)(b_I) = \kappa_J(f_i(b_I)).
$
\begin{lemma} \label{lemm: kappagi}
The function $g_i(\kappa_J)$ is simple. Further, for any $\kappa_I \in \Xi_I$ it holds that
$
\kappa_I(g_i(\kappa_J)) = (\kappa_I, \kappa_J)(f_i).
$
\end{lemma}

Proof. \quad
We know that, for any $b_I \in A_I$ it holds that
$
g_i(\kappa_J)(b_I) = \kappa_J(f_i(b_I)) = 
\sum_{k_J} c_{t_I, k_J} \cdot \kappa_J(R_J^{k_J}) \cdot \prod_{i \in I} \delta(b_i)(R_i^{t_i}),
$
where $t_i$ is such that $b_i \in R_i^{t_i}$. So,
$
g_i(\kappa_J) = \sum_{k_I} \left[ \sum_{k_J} c_{k_I, k_J} \cdot \kappa_J(R_J^{k_J}) \right] \cdot \id_{R_I^{k_I}}.
$
This is a simple function. Further, for any $\kappa_I \in A_I$ it holds that
$
\kappa_I(g_i(\kappa_J)) = \sum_{k_I} \left[ \sum_{k_J} c_{k_I, k_J} \cdot \kappa_J(R_J^{k_J}) \right] \cdot \kappa_I(R_I^{k_I}) = (\kappa_I, \kappa_J)(f_i).
$
This concludes the proof.
\hfill \qed

\vspace{-.25cm}
\subsection*{The Fubini result}

We assume that $u_i \colon A \rightarrow \re$ is a uniform limit of simple functions. 

\begin{lemma} \label{lemm: uibi exists}
Let $b_I \in A_I$. The function $u_i(b_I)$ is a uniform limit of simple functions. In particular, $u_i(b_I)$ is viable, so that $U_i(b_I)$ exists. 
\end{lemma}

Proof. \quad
Since $u_i$ is a uniform limit of simple functions, we can take a simple function $f_i$ with $f_i \le u_i \le f_i + \veps \cdot \id_A$. Then, using Lemma \ref{lemm: fibi}, $f_i(b_i)$ is a simple function with $f_i(b_I) \le u_i(b_I) \le f_i(b_I) + \veps \cdot \id_{A_J}$. It follows that $u_i(b_I)$ is a uniform limit of simple functions. In particular, $u_i(b_I)$ is viable, and $U_i(b_I)$ exists.
\hfill \qed
\vskip6pt

We can now state the Fubini result in the case of finitely additive probability measures. Let $\kappa_J \in \Xi_J$. Define $v_i(\kappa_J) \colon A_I \rightarrow \re$ by
$
v_i(\kappa_J)(b_I) = U_i(b_I)(\kappa_J)
\; \hbox{for all } b_I \in A_I.
$

\begin{theorem} \label{theo: Fubini}
{\bf (Fubini)} \quad
The function $v_i(\kappa_J)$ is a uniform limit of simple functions. Moreover, for every $\kappa_I \in \Xi_I$ it holds that:\;
$
V_i(\kappa_J)(\kappa_I) = U_i(\kappa_I, \kappa_J).
$
\end{theorem}

\noindent
Proof. \quad
Take any $\veps > 0$. Since $u_i$ is a uniform limit of simple functions, we can take a simple function $f_i$ with $f_i \le u_i \le f_i + \veps \cdot \id_A$. Take any $\kappa_J$. We show that $v_i(\kappa_J)$ is a uniform limit of simple functions. Note that $g_i(\kappa_J) \colon A_I \rightarrow \re$ is a simple function by Lemma \ref{lemm: kappagi}.
\vskip6pt

{\bf Claim.} \quad
It holds that
$
g_i(\kappa_J)(b_I) \le v_i(\kappa_J)(b_I) \le g_i(\kappa_J)(b_I) + \veps.
$

Proof of claim. \quad
Since  $f_i \le u_i \le f_i + \veps \cdot \id_A$, we know that, for every $b_I \in A_I$ and $b_J \in A_J$, it holds that
$
f_i(b_I)(b_J) \le u_i(b_I)(b_J) \le f_i(b_I)(b_J) + \veps \cdot \id_{A_J}.
$
Then, since $U_i(b_I)$ exists by Lemma \ref{lemm: uibi exists},  also
$
\kappa_J(f_i(b_I)) \le U_i(b_I)(\kappa_J) \le \kappa_J(f_i(b_I)) + \veps.
$
Then, by definition of $g_i(\kappa_J)$ and $v_i(\kappa_J)$, it holds that
$
g_i(\kappa_J)(b_I) \le v_i(\kappa_J)(b_I) \le g_i(\kappa_J)(b_I) + \veps.
$
End proof of claim.
\vskip6pt

It follows that $v_i(\kappa_J)$ is a uniform limit of simple functions. So, $v_i(\kappa_J)$ is viable, and $V_i(\kappa_J)$ exists. Take any $\kappa_I \in \Xi_I$. Since $g_i(\kappa_J)$ is a simple function with
$
g_i(\kappa_J)(b_I) \le v_i(\kappa_J)(b_I) \le g_i(\kappa_J)(b_I) + \veps
$
for every $b_I \in A_I$, it follows that
$
\kappa_I(g_i(\kappa_J)) \le V_i(\kappa_J)(\kappa_I) \le \kappa_I(g_i(\kappa_J) + \veps.
$
Further, since $f_i$ is a simple function with $f_i \le u_i \le f_i + \veps \cdot \id_A$, Lemma \ref{lemm: kappagi} implies that
$
\kappa_I(g_i(\kappa_J)) = (\kappa_I, \kappa_J)(f_i) \le U_i(\kappa_I, \kappa_J) \le (\kappa_I, \kappa_J)(f_i) + \veps = \kappa_I(g_i(\kappa_J) + \veps.
$
Hence, $V_i(\kappa_J)(\kappa_J) = U_i(\kappa_I, \kappa_J)$. This completes the proof.
\hfill \qed

\begin{lemma} \label{lemm: vi is Vi}
Let $\kappa_J \in \Xi_J$. For any $b_I \in A_I$ it holds that:\;
$
v_i(\kappa_J)(b_I) = V_i(\kappa_J)(\delta(b_I)).
$
\end{lemma}
\vskip6pt

Proof. \quad
Let $\veps > 0$. Since $v_i(\kappa_J)$ is a uniform limit of simple functions, we can take a simple function $f_i \colon A_I \rightarrow \re$ with
$
f_i(b_I) \le v_i(\kappa_J)(b_I) \le f_i(b_I) + \veps
$
for all $b_I \in A_I$. Then, for all $b_I \in A_I$,
$
f_i(b_I) = \delta(b_I)(f_i) \le V_i(\kappa_J)(\delta(b_I)) \le \delta(b_I)(f_i) + \veps = f_i(b_I) + \veps.
$
It follows that $v_i(\kappa_J)(b_I) = V_i(\kappa_J)(\delta(b_I))$ for all $b_I \in A_I$.
\hfill \qed

\vspace{-.25cm}
\section*{Appendix D. \quad Detailed Proofs for Section \ref{Section: Admissibility}}

\begin{lemma} \label{theo: vi continuous}
Suppose that $u_i$ is continuous and a uniform limit of simple functions. Then the function $v_i(\kappa_J)$ is continuous.
\end{lemma}

Proof. \quad
Take any $\kappa_J \in \Sigma_J$ and any $b_I \in A_I$. We show that $v_i(\kappa_J)$ is continuous at $b_I$. Let $\veps > 0$. We show that there is an open set $U \ni b_I$ with
$
\vert v_i(\kappa_J)(s_I) - v_i(\kappa_J)(b_I) \vert \le 4 
\cdot \veps
$
for all $s_I \in U$.
\vskip6pt

Since $u_i$ is a limit of simple functions, we can take a simple function 
$
f_i = \sum_{k \in K} c_k \cdot \id_{R^k}
$
in product type representation with $f_i \le u_i \le f_i + \veps$.
\vskip6pt

For $s_i \in A_i$ and $k \in K_i$, we say that $R_i^k$ is adjacent to $s_i$ if for every open set $V_i \ni s_i$ it holds that $V_i \cap R_i^k \not= \emptyset$.
\vskip6pt

For each $i$, let $L_i$ be the set of indices $k \in K_i$ for which $R^k_i$ is adjacent to $b_i$. For each index $k \notin L_i$, let $V_i^k \ni b_i$ be such that $V_i^k \cap R_i^k = \emptyset$. Write
$
U_i = \bigcap_{k \notin L_i} V_i^k
\quad \hbox{and} \quad
U_I = \prod_{i \in I} U_i.
$
\vskip6pt

{\bf Claim 1.} \quad
It holds that
$
f_i(b_I)(s_J) - 2 \cdot \veps \le u_i(s_I)(s_J) \le f_i(b_I)(s_J) + 2 \cdot \veps
$
for every $s_I \in U_I$ and $s_J \in A_J$.
\vskip6pt

Proof. \quad
Let $s_I \in U_I$ and $s_J \in A_J$. Let $\ell_j$ be the index in $K_j$ for which $s_j \in R^{\ell_j}_j$.
\vskip6pt

Let $k_I \in L_I$ with $s_I \in R^{k_I}_i$. Then, since $u_i$ is continuous, and $R^{k_I}_I$ is adjacent to $b_I$, there is $s^{k_I}_I \in U_I$ with $(s^{k_I}_I, s_J) \in R_I^{k_I} \times R^{\ell_J}_J$, and
$
\vert u_i(b_I, s_J) - u_i(s^{k_I}_I, s_J) \vert < \veps.
$
Further, since both $(s_I, s_J) \in R_I^{k_I} \times R^{\ell_J}_J$ and $(s_I^{k_I}, s_J) \in R_I^{k_I} \times R^{\ell_J}_J$, it holds that
$
u_i(s_I, s_J) \ge f_i(s_I, s_J) = f_i(s_I^{k_I}, s_J) \ge u_i(s_I^{k_I}, s_J) - \veps.
$
Then
$$
u_i(s_I, s_J) \ge u_i(s^k_I, s^k_J) - \veps \ge u_i(b_I, s^k_J) - 2 \cdot \veps \ge f_i(b_I)(s^k_J) - 2 \cdot \veps = f_i(b_I)(s_J) - 2 \cdot \veps.
$$
We find that
$
f_i(b_I)(s_J) + 2 \cdot \veps \ge u_i(s_I, s_J) \ge f_i(b_I)(s_J) - 2 \cdot \veps.
$
This completes the proof of Claim 1.
\vskip6pt

So, we know now that
$
\kappa_J(f_i(b_I)) + 2 \cdot \veps \ge U_i(s_I)(\kappa_J) \ge \kappa_J(f_i(b_I)) - 2 \cdot \veps.
$
Since this is true for any $s_I \in U_I$, and since in particular $b_I \in U_I$, it follows that
$
\vert U_i(s_I)(\kappa_J) - U_i(b_I)(\kappa_J) \vert \le 4 \cdot \veps
$
for all $s_I \in U$. Hence, by definition of $v_i$, it follows that
$
\vert v_i(\kappa_J)(s_I) - v_i(\kappa_J)(b_I) \vert \le 4 \cdot \veps.
$
This completes the proof.
\hfill \qed

\begin{lemma} \label{theo: undominated BRs}
Let $\Gamma = (N, A, \F, \Sigma, u)$ be a friendly game. Let $\tau \in \Xi$ be such that $\tau_j$ is strictly positive for every $j \not= i$. If $\rho_i$ is dominated by $\mu_i$, then
$
U_i(\rho_i \mid \tau) < U_i(\mu_i \mid \tau).
$
\end{lemma}

\noindent
Proof.\quad
Since $\mu_i$ dominates $\rho_i$, we know that
$
U_i(\rho_i, \delta(b_{-i})) \le U_i(\mu_i, \delta(b_{-i}))
$
for all $b_{-i} \in A_{-i}$. We show that there is a $b_{-i} \in A_{-i}$ with
$
U_i(\rho_i, \delta(b_{-i})) < U_i(\mu_i, \delta(b_{-i})).
$
Suppose by contradiction that $U_i(\rho_i, \delta(b_{-i})) = U_i(\mu_i, \delta(b_{-i}))$ for all $b_{-i} \in A_{-i}$. Then by Lemma \ref{lemm: vi is Vi} and Theorem \ref{theo: Fubini},
\begin{eqnarray*}
v_i(\rho_i)(b_{-i}) = V_i(\rho_i)(\delta(b_{-i}))
& = &
U_i(\rho_i, \delta(b_{-i})) \cr
& = &
U_i(\mu_i, \delta(b_{-i})) = V_i(\mu_i)(\delta(b_{-i})) = v_i(\mu_i)(b_{-i})
\end{eqnarray*}
for any $b_{-i} \in A_{-i}$. So, $v_i(\rho_i) = v_i(\mu_i)$. Then, using Theorem \ref{theo: Fubini},
$
U_i(\rho_i, \tau_{-i}) = V_i(\rho_i)(\tau_{-i}) = V_i(\mu_i)(\tau_{-i}) = U_i(\mu_i, \tau_{-i})
$
for any profile $\tau_{-i}$. This contradicts the assumption that $\mu_i$ dominates $\rho_i$. So, there is at least one $b_{-i} \in A_{-i}$ with
$
U_i(\rho_i, \delta(b_{-i})) < U_i(\mu_i, \delta(b_{-i})).
$
Then, since $v_i(\rho_i)$ and $v_i(\mu_i)$ are continuous by Lemma \ref{theo: vi continuous}, there are open sets $Z_j$ for $j \not= i$ with
$
v_i(\rho_i)(b_{-i}) = U_i(\rho_i, \delta(b_{-i})) < U_i(\mu_i, \delta(b_{-i})) = v_i(\mu_i)(b_{-i})
$
for all $b_{-i} \in Z_{-i}$. Then, since $\tau_j(Z_j) > 0$ for all $j \not= i$, it follows that 
$
V_i(\rho_i)(\tau_{-i}) < V_i(\mu_i)(\tau_{-i})
$
Hence, indeed
$
U_i(\rho_i \mid \tau) = V_i(\rho_i)(\tau_{-i}) < V_i(\mu_i)(\tau_{-i}) = U_i(\mu_i \mid \tau).
$
\hfill \qed

\vspace{-.25cm}
\section*{Appendix E. \quad Detailed Proofs for Section \ref{Section: Invariance}}

\begin{lemma} \label{lemm: Phi surjectie}
For every $\eta \in \Xi(B)$ there is $\zeta \in \Xi(A)$ with $\Phi(\zeta) = \eta$.
\end{lemma}

Proof. \quad
Take any $\eta \in \Xi(B)$. Define
$
\H = \{ \Phi^{-1}(G) \mid G \in \G \}.
$
Then $\H$ is a subfield of $\F$. For each $H \in \H$, define $\zeta(H) = \eta(\Phi(H))$. Then $\zeta$ is a finitely additive strategy profile on $\H$.
The result now follows from Theorem 2 in \cite{Los Marczewski: 1949}. See also Lemma C.2. in \cite {Zseleva: 2017}.
\hfill \qed
\vskip6pt

{\bf Proof of Claim A.} \quad
Suppose that $\kappa = c \cdot \eta + (1-c) \cdot \xi$ for some $c \in (0, 1)$. Let $E_1$, $E_2$, $E_3$ be a partition of $\na$. Then $\eta(E_j) = 1$ for precisely one $j$, and the same holds for $\xi$. Hence, $\kappa(E_j) = 0$ for at least one $j$.
\vskip6pt

Conversely, suppose that for any partition $E_1, E_2, E_3$ of $\na$ it holds that $\kappa(E_i) = 0$ for at least one $i = 1, 2, 3$. If for every set $E \subseteq \na$ we have that either $\kappa(E) = 0$ or $\kappa(\na \setminus E) = 0$. Then $\kappa$ is an ultrafilter. Hence, $\kappa$ is a hazy filter.
\vskip6pt

So, suppose that there is a set $E \subseteq \na$ with $\kappa(E) > 0$ and $\kappa(\na \setminus E) > 0$. Write $c = \kappa(E)$. Write $\eta(K) = \frac{\kappa(K \cap E)}{c}$, and $\xi(K) = \frac{\kappa(K \cap (\na \setminus E))}{1-c}$. Then $\kappa = c \cdot \eta + (1-c) \cdot \xi$, and $\eta(E) = 1$ and $\xi(\na \setminus E) = 1$.
\vskip6pt

We show that $\eta$ is an ultrafilter. Consequently, also $\xi$ is an ultrafilter, which completes the proof. Take any set $X \subseteq E$. Write $E_1 = \na \setminus E$, $E_2 = E \setminus X$, and $E_3 = X$. Then by [2], we have $\kappa(E_m) = 0$ for at least one $E_m$.
Since $\kappa(\na \setminus E) > 0$, it follows that $\kappa(X) = 0$ or $\kappa(E \setminus X) = 0$.
Hence, for any set $X \subseteq E$ it holds that $\eta(X) = 0$ or $\eta(E \setminus X) = 0$. Hence, since $\eta(E) = 1$, $\eta$ is a ultrafilter.
\hfill \qed
\vskip6pt

{\bf Proof of Claim B in Example \ref{example: hazy filters}.} \quad
We prove Claim B in three Steps. We first need to specify the pure best responses of player 2 against a strategy of player 1 with maximal carrier.\footnote{A strategy $(p, 1-p)$ has a maximal carrier if $0 < p < 1$.}
Note that, when $\kappa_1 = (p, 1-p)$ with $0 < p < 1$, and player 2 plays $k \ge 2$, $k \not= \infty$, then
$
U^k := U_2(\kappa_1, \delta(k)) = \frac{1}{k} \cdot \left( p - \frac{1-p}{k} \right).
$
For $k \ge 2$, $k \not= \infty$, define
$
p_k = \frac{2k+1}{k^2+3k+1}.
$
\vskip6pt

{\bf Step 1.} \quad
It holds that $1 > p_2 > p_3 > p_4 > \cdots \cdots > 0$. Further, for each $p_k$, the best responses for player 2 are $k$ and $k+1$. For $p$ with $p_k > p > p_{k+1}$, the unique best response for player 2 is $k+1$.
\vskip6pt

Proof of Step 1. \;
Note that $U^k \le U^{k+1}$ when
$
\frac{1}{k} \cdot \left( p - \frac{1-p}{k} \right) \le \frac{1}{k+1} \cdot \left( p - \frac{1-p}{k+1} \right).
$
Solving for $p$ yields
$
p \le \frac{2k+1}{k^2+3k+1} = p_k.
$
The statement now follows from the observation that $p_k$ is strictly decreasing in $k$, and that $u^k > 0$, while playing action $\infty$ yields payoff zero. End proof of Step 1.
\hfill \qed
\vskip6pt

{\bf Step 2.} \quad
Let $(\kappa_1, \kappa_2)$ be a perfect equilibrium of the game $\Gamma$. Then $\kappa_1 = \delta(D)$ and $\kappa_2$ is a twinned hazy filter.
\vskip6pt

Proof of Step 2. \quad
Let $\kappa = (\kappa_1, \kappa_2)$ be any finitely additive perfect equilibrium of $\Gamma$. 
Since $D$ is the unique dominant strategy for player 1, $\kappa_1 = \delta(D)$ by Theorem \ref{theo: limit undominated}. Further, for any $k = 1, 2, \ldots, \infty$ it follows from Step 1 that $k$ is not a best response against any maximal carrier strategy of player 1 sufficiently close to $D$. So, $\kappa_2(k) = 0$ for any $k = 1, 2, \ldots, \infty$. Hence, $\kappa_2$ is diffuse.
\vskip6pt

We show that $\kappa$ is a hazy filter.
Let $\sigma_\alpha = (p_\alpha, \sigma_{\alpha, 2})$ and $\kappa_\alpha = ((0, 1), \kappa_{\alpha,2})$ be as in Definition \ref{def-perfstr} of perfect equilibrium. Since $\kappa_{\alpha,2}$ is a best response against the maximal carrier strategy $p_\alpha$, it follows from Step 1 that $\kappa_{\alpha, 2}(E_i) = 0$ for at least one $i = 1, 2, 3$. Since $\kappa_{\alpha, 2} \rightarrow \kappa_2$, this then also holds for $\kappa_2$. So, $\kappa_2$ is a hazy filter by Claim A.
\vskip6pt

So, we can write $\kappa_2 = c \cdot \eta + (1-c) \cdot \xi$ for some $c \in (0, 1)$ and ultrafilters $\eta$ and $\xi$.
We show that $\xi$ and $\eta$ are twins.
\vskip6pt

If $\eta = \xi$, it is clear that $\eta$ and $\xi$ are twins. So, assume that $\eta \not= \xi$. Then there is a set $E \subseteq \na$ with $\eta(E) = 1$ and $\xi(\na \setminus E) = 1$.
Take any two subsets $X$ and $Y$ of $\na$ with $\xi(X) = \eta(Y) = 1$. We argue that there is a set $B_k = \{ k, k+1 \}$ with both $X \cap B_k \not= \emptyset$ and $Y \cap B_k \not= \emptyset$. We can assume wlog that $X \subseteq E$ and $Y \subseteq \na \setminus E$. In particular, $X$ and $Y$ are disjoint.
\vskip6pt

Note that $\kappa_2(X) > 0$ and $\kappa_2(Y) > 0$. So, since $\kappa_{\alpha,2} \rightarrow \kappa_2$, we know that, for any $\alpha$ sufficiently far into the net, $\kappa_{\alpha,2}(X) > 0$ and $\kappa_{\alpha,2}(Y) > 0$. Take such an $\alpha$. Then, according to Step 1, we know that $p_\alpha = p_k$ for some $k \in \na$. Then $X \cap B_k \not= \emptyset$, and $Y \cap B_k \not= \emptyset$. End proof of Step 2.
\hfill $\triangleleft$
\vskip6pt

{\bf Step 3.} \quad
Let $\kappa = (\kappa_1, \kappa_2)$ be a strategy pair with $\kappa_1 = \delta(D)$ and $\kappa_2$ is a twinned hazy filter. Then $\kappa$ is a perfect equilibrium of $\Gamma$.
\vskip6pt

Proof of Step 3. \quad
Let $\kappa_2$ be a twinned hazy filter. So, there are twins $\eta$ and $\xi$ with
$
\kappa_2 = c \cdot \eta + (1 - c) \cdot \xi
$
for some $c \in (0, 1)$. Let $F = (F_1, \ldots, F_m)$ be a partition of $\na$. The proof is divided in two parts.
\vskip6pt

{\bf Part A.} \quad
There are $a, b \in \na$ with $|a-b| \le 1$ such that
$$
\delta(a)(F_\ell) = \eta(F_\ell)
\quad \hbox{ and } \quad
\delta(b)(F_\ell) = \xi(F_\ell)
$$
for every $\ell = 1, \ldots, m$.
\vskip6pt

Proof of Part A. \quad
Since $\eta$ and $\xi$ are ultrafilters, there are $F_s$ and $F_t$ with $\eta(F_s) = \xi(F_t) = 1$. If $F_s = F_t$, then we take $a = b \in F_S$. So, assume that $F_s \not= F_t$. Since $\eta$ and $\xi$ are twins, there is $k$ with $F_s \cap B_k \not= \emptyset$, and $F_t \cap B_k \not= \emptyset$. Write $F_s \cap B_k = \{ a \}$ and $F_t \cap B_k = \{ b \}$. Then $a$ and $b$ are as required. End proof of Part A.
\vskip6pt

{\bf Part B.} \quad
We complete the proof of Step 3. Let $I$ be the collection of partitions $F = (F_1, \ldots, F_m)$ of $\na$, where $m \in \na$ is arbitrary. For $F \in I$ and $\veps > 0$, write
$$
U(F, \veps) = \left\{ \rho \in \Xi_2 \mid 
\left| \kappa_2(F_i) - \rho(F_i) \right| < \veps\hbox{ for every } i = 1, \ldots m \right\}.
$$
Note that for every open set $U \ni \kappa_2$ there are $F \in I$ and $\veps > 0$ with $U(F, \veps) \subseteq U$.
So, by item [2] of Theorem \ref{prop: characterization} it is sufficient to show that, for every $F$ and every $\veps > 0$, there are $(p, 1-p)$ with $0 < p < \veps$ and $\rho \in U(F, \veps)$ such that $\rho$ is a best response to $(p, 1-p)$.
\vskip6pt

First, choose $K \in \na$ such that $p_K < \veps$. We may assume that
$$
F_i \subseteq \{ K, K+1, \ldots \} \quad \hbox{ for all infinite } F_i.
$$
If this is not the case, we can switch to a refinement of $F$.
Then, given $F = (F_1, \ldots, F_m)$, choose $a$ and $b$ as in Part A. Define
$$
\rho = c \cdot \delta(a) + (1-c) \cdot \delta(b).
$$
Then $\rho(F_i) = \kappa_2(F_i)$ for every $i = 1, \ldots, m$, so that indeed $\rho \in U(F, \veps)$. Choose $k$ such that $\{a, b \} \subseteq \{k, k+1\}$. Then $k \ge K$, so that $p_k < \veps$.
Then, by Step 1, $\rho$ is a best response to $p_k$.
End proof of Step 3.
\hfill \qed

\vspace{-.25cm}
\section*{\large Appendix F. \quad Detailed Proof of Theorem \ref{theorem-mainexistence}.}

Our main existence result uses the fixed point Theorem of Kakutani-Fan-Glicksberg. For reference, we formally state this Theorem.

\begin{theorem}[Kakutani-Fan-Glicksberg] \label{theo: Glicksberg}
Let $K$ be a non-empty, convex, and compact subset of a locally convex topological vector space. Let $\varphi \colon K \dra K$ be a correspondence whose graph is closed, and whose values are non-empty and convex.
Then, the set of fixed points of $\varphi$ is compact and non-empty.
\end{theorem}

\noindent
Proof. \quad
Step {\bf A.} \quad
Take any strategy profile $\tau \in \Sigma$, and any $\veps \in (0, 1)$.
In this step, we define a correspondence $\BR(\tau, \veps)$, and show that it has at least one fixed point. For each player $i\in N$, define the set $\Sigma_i(\tau_i, \veps) \subseteq \Sigma_i$ by 
$$
\Sigma_i(\tau_i, \veps) =
\Bigl\{ (1 - \veps) \cdot \sigma_i + \veps \cdot \tau_i \mid \sigma_i \in \Sigma_i \Bigr\}.
$$
Define $\Sigma(\tau, \veps) = \prod_{i\in N} \Sigma_i(\tau_i, \veps)$.
For each player $i$, define the best-response correspondence $\BR(\tau, \veps)_i \colon \Sigma(\tau, \veps) \dra \Sigma_i(\tau_i, \veps)$ by, for every $\sigma \in \Sigma(\tau, \veps)$,
$$
\BR(\tau, \veps)_i(\sigma) = \Bigl\{ \kappa_i \in \Sigma_i(\tau_i, \veps) \mid U_i(\kappa_i \mid \sigma) \ge U_i(\rho_i \mid \sigma) \text{ for all } \rho_i \in \Sigma_i(\tau_i, \veps) \Bigr\}.
$$
The set $\BR(\tau, \veps)_i(\sigma)$ is the collection of best responses of payer $i$ to the strategy profile $\sigma$ within the set $\Sigma_i(\tau_i, \veps)$. Define $\BR(\tau, \veps) \colon \Sigma(\tau, \veps) \dra \Sigma(\tau, \veps)$ by $\BR(\tau, \veps)(\sigma) = \prod_{i\in N} \BR(\tau, \veps)_i(\sigma)$. We show that $K = \Sigma(\tau, \veps)$ and $\varphi = \BR(\tau, \veps)$ satisfy the conditions of Theorem \ref{theo: Glicksberg}. Non-emptiness and compactness of $\Sigma(\tau, \veps)$ follow from the fact that this set is the image of the non-empty and compact set $\Sigma$ under the continuous map
$$
\sigma \mapsto (1 - \veps) \cdot \sigma + \veps \cdot \tau.
$$
Continuity of this map follows from the continuity of addition and scalar multiplication. Convexity of $\Sigma(\tau, \veps)$ follows from the linearity of this map and the convexity of $\Sigma$.
\vskip6pt

Closedness of the graph of  $\BR(\tau, \veps)$ is a consequence of the continuity of the payoff functions $U_i$.
Convex-valuedness of $\BR(\tau, \veps)$ follows from the multi-linearity of the payoff functions $U_i$. Non-emptiness follows from non-emptiness and compactness of $\Sigma(\tau, \veps)$, and continuity of the functions $U_i$. Hence, by Theorem \ref{theo: Glicksberg}, for any choice of $\tau$ and $\veps$ the correspondence $\BR(\tau, \veps)$ has at least one fixed point.
\vskip6pt

Step {\bf B.} \quad
Take any carrier $\C$. In this step we show that $\text{PS}(\C)$ is compact and non-empty. Compactness of $\hbox{PS}(\C)$ follows from the observation that $\hbox{PS}(\C)$ is a closed subset of the compact space $\Sigma$. We prove non-emptiness. For each $i \in N$, take $\tau_i \in \Sigma_i$ with $\C_i(\tau_i) = \C_i$. For each $k \in \na$, let $\sigma^k$ be a fixed point of $\BR(\tau, \frac{1}{k})$, which exists by Step A. 
Since $(\sigma^k)_{k \in \na}$ is a sequence in the compact space $\Sigma$, there is a subnet $(\sigma_{\phi(\beta)})_{\beta \in J}$, with associated map $\phi \colon J \rightarrow \na$, that converges to some strategy profile $\sigma \in \Sigma$.\footnote{For a detailed discussion of the existence of such a subnet, see for example Theorem 2.31 of \cite{Aliprantis: 2006}.} We show that $\sigma \in \hbox{PS}(\C)$.
\vskip6pt

It suffices to construct a net $(\kappa_{\phi(\beta)})_{\beta \in J}$ that also converges to $\sigma$, while for each $\beta \in J$ $\kappa_{\phi(\beta)}$ is a best response to $\sigma_{\phi(\beta)}$. Take any $\beta \in J$. Write $k = \phi(\beta)$. Since $\sigma^k \in \Sigma(\tau, \frac{1}{k})$, we have 
\begin{equation}\label{eq-1stconvexcomb}
\sigma^k = \frac{k-1}{k} \cdot \kappa^k + \frac{1}{k} \cdot \tau
\end{equation}
for some $\kappa^k \in \Sigma$. We define $\kappa_{\phi(\beta)} = \kappa^k$.
\vskip6pt

We show that the net $(\kappa_{\phi(\beta)})_{\beta \in J}$ converges to $\sigma$, and that each $\kappa_{\phi(\beta)}$ is a best response to $\sigma_{\phi(\beta)}$. By equation \ref{eq-1stconvexcomb} we have
$
\kappa_{\phi(\beta)} = \frac{k}{k - 1} \cdot \sigma_{\phi(\beta)} - \frac{1}{k-1} \cdot \tau.
$
Hence, the net $(\kappa_{\phi(\beta)})_{\beta \in J}$ converges to $\sigma$ by continuity of addition and scalar multiplication.
\vskip6pt

It remains to show that each $\kappa_{\phi(\beta)}$ is a best response to $\sigma_{\phi(\beta)}$. Take any $\beta \in J$. Write $k = \phi(\beta)$. Fix a player $i\in N$ and take any strategy $\zeta_i \in \Sigma_i$. We show that $U_i(\zeta_i \mid \sigma^k) \leq U_i(\kappa_i^k \mid \sigma^k)$. Consider 
\begin{equation}\label{eq-2ndconvexcomb}
\rho_i^k = \frac{k-1}{k} \cdot \zeta_i + \frac{1}{k} \cdot \tau_i.
\end{equation}
Note that $\rho_i^k \in \Sigma_i(\tau_i, \frac{1}{k})$. Because $\sigma^k$ is a fixed point of $\BR(\tau, \frac{1}{k})$, we have 
$
U_i(\rho_i^k \mid \sigma^k) \leq U_i(\sigma_i^k \mid \sigma^k).
$
Hence, by Eqs.~\eqref{eq-1stconvexcomb} and \eqref{eq-2ndconvexcomb} and by multi-linearity of $U_i$, we obtain
$
U_i(\zeta_i \mid \sigma^k) \,\leq\, U_i(\kappa_i^k \mid \sigma^k),
$
as desired. So, $\sigma \in \hbox{PS}(\C)$. 
\vskip6pt

\noindent
Step {\bf C.} \quad
We show that $\hbox{PS}$ is non-empty, and compact. We need the following terminology and observations.
A collection of sets $(K_i)_{i \in I}$ has the finite intersection property, if for every non-empty finite subset $F$ of $I$, $\bigcap_{I \in F} K_i$ is non-empty.
It is known that a topological space is compact if and only if every collection of closed subsets having the finite intersection property has a non-empty intersection.\footnote{This is simply the contraposition of the standard definition of compactness. For a reference, see for example \cite{Aliprantis: 2006}, p.39, Theorem 2.31.}
\vskip6pt

Thus, since $\Sigma$ is compact, by Step B it suffices to show that the collection
$
{\mathcal P} = \{\text{PS}(\C)\mid \C\text{ is a carrier}\}
$
has the finite intersection property. Note that, for any carriers $\C_1,\ldots,\C_k$, the carrier of a strictly convex combination of strategy profiles $\sigma^\ell$ with $\C(\sigma^\ell) = \C_\ell$ equals the union of the carriers of the respective strategy profiles $\sigma^\ell$. So,
$
\D = \bigcup_{j=1,\ldots,k}\C_j
$
is also a carrier. Moreover, as is shown in Step B, the set PS$(\D)$ is non-empty. It follows that
$\bigcap_{j=1,\ldots,k}\text{PS}(\C_j)\,\supseteq\,\text{PS}( \D )\,\neq\,\emptyset.$
Hence, ${\mathcal P}$ has the finite intersection property. 
Thus PS is non-empty, and compact. Theorem \ref{theorem-mainexistence} now follows directly from Theorem \ref{theo: perfect eq}.
\hfill \qed

\end{document}